\documentclass{article} % For LaTeX2e
\usepackage{iclr2026_conference,times}
\usepackage{graphicx}

\usepackage{amsmath,amsfonts,bm}

\def\eqref#1{equation~\ref{#1}}
\def\1{\bm{1}}

\DeclareMathAlphabet{\mathsfit}{\encodingdefault}{\sfdefault}{m}{sl}
\SetMathAlphabet{\mathsfit}{bold}{\encodingdefault}{\sfdefault}{bx}{n}

\usepackage{hyperref}
\usepackage{url}
\usepackage{subcaption}
\usepackage{float}

\title{Factor Dimensionality and the Bias–Variance Tradeoff in Diffusion Portfolio Models}

\author{Avi Bagchi\thanks{Equal contribution.} \quad
Michael Tesfaye\footnotemark[1] \quad
Om Shastri \\
University of Pennsylvania \\
Philadelphia, PA 19104, USA \\
\texttt{\{aviba,tesfaye,oshastri\}@seas.upenn.edu}
}

\iclrfinalcopy

\track{Industry \& Applications}
\begin{document}

\maketitle

\begin{abstract}
In this paper, we implement and evaluate a conditional diffusion model for asset return prediction and portfolio construction on large-scale equity data. Our method models the full distribution of future returns conditioned on firm characteristics (i.e.\ factors), using the resulting conditional moments to construct portfolios. We observe a clear bias--variance tradeoff: models conditioned on too few factors underfit and produce overly diversified portfolios, while models conditioned on too many factors overfit, resulting in unstable and highly concentrated allocations with poor out-of-sample performance. Through an ablation over factor dimensionality, we reveal an intermediate number of factors that achieves the best generalization and outperforms baseline portfolio strategies. 

% We also motivate methods where low-dimensional structure is learned implicitly during score estimation, without explicit factor selection.

\end{abstract}

% \textbf{Second}, we propose the framework of \citet{chen2026diffusionfactormodelsgenerating, chen2023scoreapproximationestimationdistribution} as an alternative to costly factor ablation, where the diffusion process implicitly learns a low-dimensional structure during score estimation, adaptively balancing this tradeoff during training.

\section{Introduction}

Predicting asset returns is a fundamental problem in quantitative finance. Linear factor models \citep{fama1993common, fama2015five} provide a tractable framework for modeling asset returns but struggle to capture nonlinear and higher-order market dynamics. \citet{chen2026diffusionfactormodelsgenerating} introduces generative approaches that learn full conditional return distributions rather than point forecasts. In this paper, we evaluate the conditional diffusion framework of \citet{gao2025factorbasedconditionaldiffusionmodel}, which generates returns conditioned on observable firm characteristics (i.e. factors). We show that factor dimensionality induces a clear bias–variance tradeoff in diffusion-based return modeling: too few factors lead to underfitting and an excessively diverse portfolio, while too many produce high-variance models with overly concentrated allocations. Empirical ablations reveal an optimal dimensionality that outperforms baseline strategies. We use data from Wharton Research Data Services (WRDS) based on the procedure specified by \citet{JensenKellyPedersen2023}. We defer dataset details and related work to the appendix (Appendix~\ref{app:data}, Appendix~\ref{app:related_work}).

\section{Diffusion-Based Conditional Return Modeling}
We follow \citet{gao2025factorbasedconditionaldiffusionmodel} which formulates asset return prediction as learning a conditional return distribution given observable firm characteristics. $R_{t+1} \in \mathbb{R}^N$ denotes a vector of returns for $N$ assets observed in time periods $t = 1 \dots T$ and let $X_t = \{X_{i,j}\}^N_{i=1}$ denote the corresponding set of asset-level characteristics (i.e. factors) observed at time $t$. Returns are assumed to satisfy $R_{t+1} = f(X_t) + \epsilon_{t+1}$,\ where $f(\cdot)$ is an unknown, potentially nonlinear function and $\epsilon_{t+1}$ captures unpredictable shocks independent of information known at $t$. The objective is to learn the full conditional distribution $p(R_{t+1} | X_t)$, rather than only conditional means.

To estimate this distribution, we adopt a conditioning denoising diffusion probabilistic model \citep{ho2020denoisingdiffusionprobabilisticmodels}. The forward diffusion process gradually corrupts observed returns by adding Gaussian noise over a fixed number of steps, transforming the data into an isotropic Gaussian distribution. A neural network is then trained to reverse this process by predicting the noise added at each diffusion step, conditional on characteristics $X_t$. The reverse diffusion process is implemented using a diffusion transformer architecture. Following \citet{gao2025factorbasedconditionaldiffusionmodel} in modifying \citet{peebles2023scalablediffusionmodelstransformers}, each asset is represented as a token and cross-sectional dependence among assets is captured through self-attention layers. Conditioning on firm characteristics is performed locally at the token level via adaptive normalization layers. This approach allows the denoising dynamics of each asset to depend on its own characteristics while still modeling joint return behavior across assets. After training, the model generates Monte Carlo samples from the conditional distribution $p(R_{t+1} | x_t)$ for each period, which are used to estimate the conditional mean and covariance of returns that serve as inputs to the portfolio construction procedure (i.e. mean–variance optimization).

\section{Results}

Each month $t$, we estimate the conditional mean vector $\hat{\mu_t}$ and the covariance matrix $\hat{\Sigma}_t$ of the next-month returns. We then compute long-only portfolio weights solving a constrained mean-variance optimization problem $\max \;\;\omega^\top\hat{\mu_t}-\frac{\gamma}{2}\omega^\top \hat{\Sigma}_t\omega \;\;\text{subject to}\;\; 1^\top\omega = 1 \;\; \text{and} \;\; \omega \geq 0$ with $\mu$ as the expected return vector, $\Sigma$ as the return covariance matrix, $\gamma=100$ as the risk-aversion parameter, and $\omega$ are portfolio weights \citep{markowitz1952portfolio}. We follow \citet{gao2025factorbasedconditionaldiffusionmodel} in comparing the diffusion factor portfolio with three simpler baseline portfolios (Appendix~\ref{app:portfolio_constructions}).

\begin{figure}[t]
\centering

\begin{subfigure}[t]{0.32\linewidth}
    \centering
    \includegraphics[width=\linewidth]{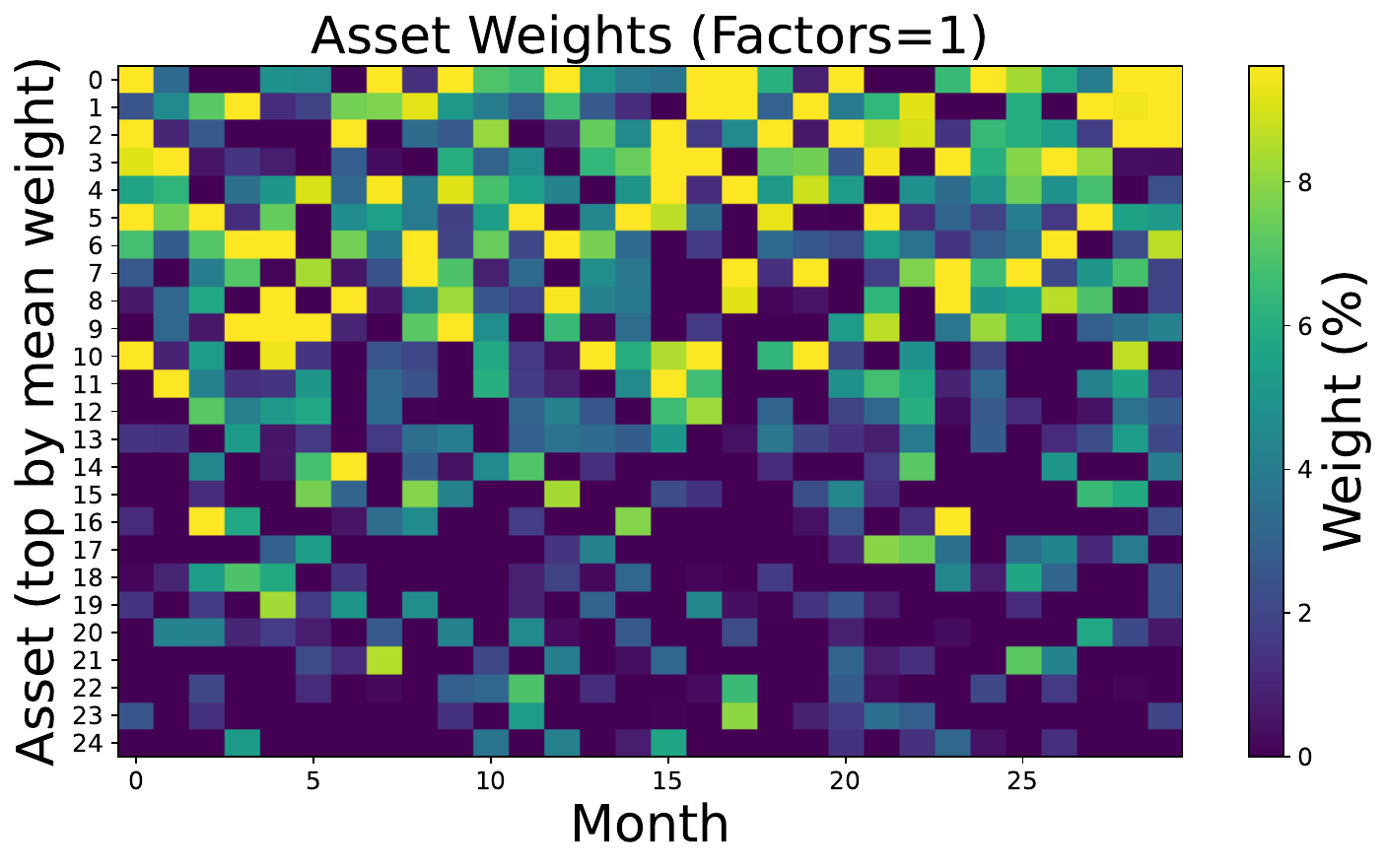}
    \caption{Low capacity (high bias)}
\end{subfigure}
\hfill
\begin{subfigure}[t]{0.32\linewidth}
    \centering
    \includegraphics[width=\linewidth]{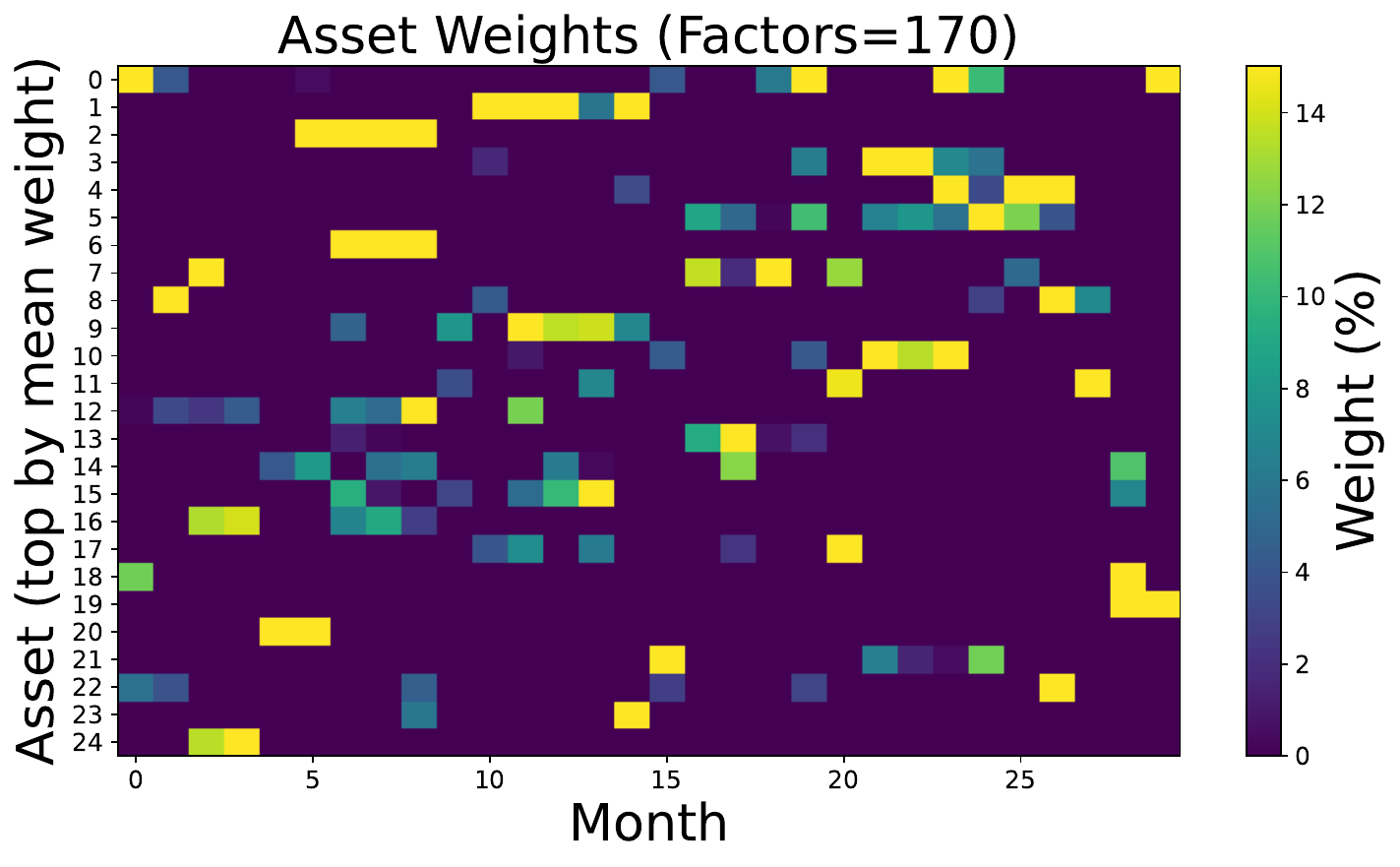}
    \caption{Medium capacity}
\end{subfigure}
\hfill
\begin{subfigure}[t]{0.32\linewidth}
    \centering
    \includegraphics[width=\linewidth]{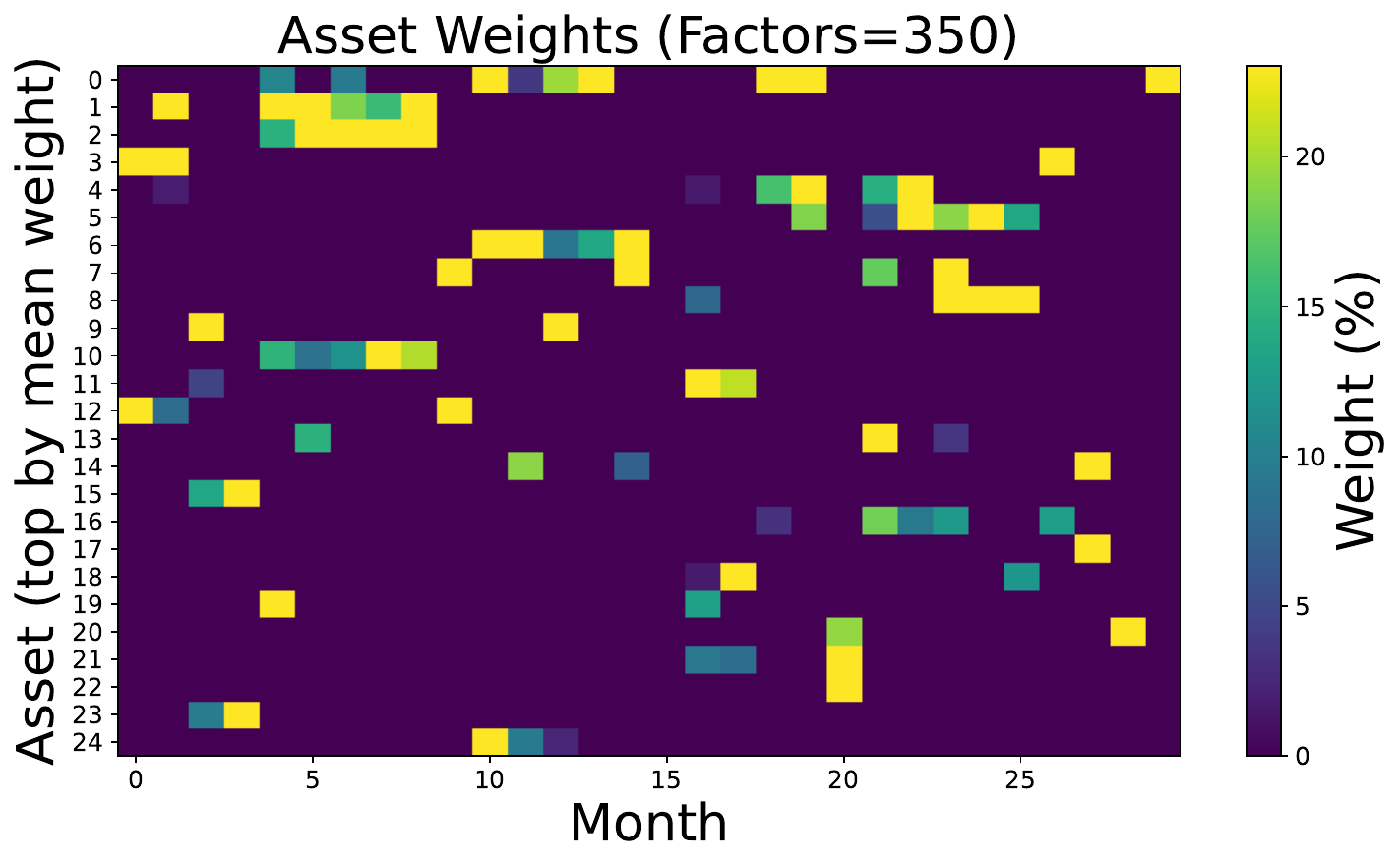}
    \caption{High capacity (high variance)}
\end{subfigure}

\caption{Heatmaps for 200 samples show monthly asset weights (top 25 assets by average allocation over the time period) learned under different factor dimensionalities. A low-capacity model (left) distributes weight broadly, reflecting underfitting and high bias. An intermediate model (middle) concentrates allocations on persistent signals, indicating effective factor utilization. A high-capacity model (right) produces sparse, unstable allocations consistent with overfitting and high variance.}
\label{fig:three_panel_weights}
\end{figure}

For small $k$, the portfolio weights are relatively dispersed across assets, reflecting a low-capacity model that produces broadly diversified allocations (Figure~\ref{fig:three_panel_weights}). As $k$ increases, the weight distribution becomes more concentrated, with larger positions placed on a smaller set of assets (Figure~\ref{fig:three_panel_weights}). We observe that the moderately diverse portfolio (b) outperforms EW, Emp, and ShrEmp in terms of cumulative returns, where the low capacity (a) ($k=1$) and the high capacity (c) ($k=350$) fail to do so (Figure~\ref{fig:three_panel_returns}). See Appendix~\ref{app:cumm_ret} and Appendix~\ref{app:port_weight} for the full ablation results. Future work should evaluate these results against the framework of \citet{chen2026diffusionfactormodelsgenerating, chen2023scoreapproximationestimationdistribution} which implicitly learns a low-dimensional factor structure through score decomposition during score estimation, eliminating the need for explicit factor selection (Appendix~\ref{app:implicit}).

\begin{figure}[t]
\centering

\begin{subfigure}[t]{0.32\linewidth}
    \centering
    \includegraphics[width=\linewidth]{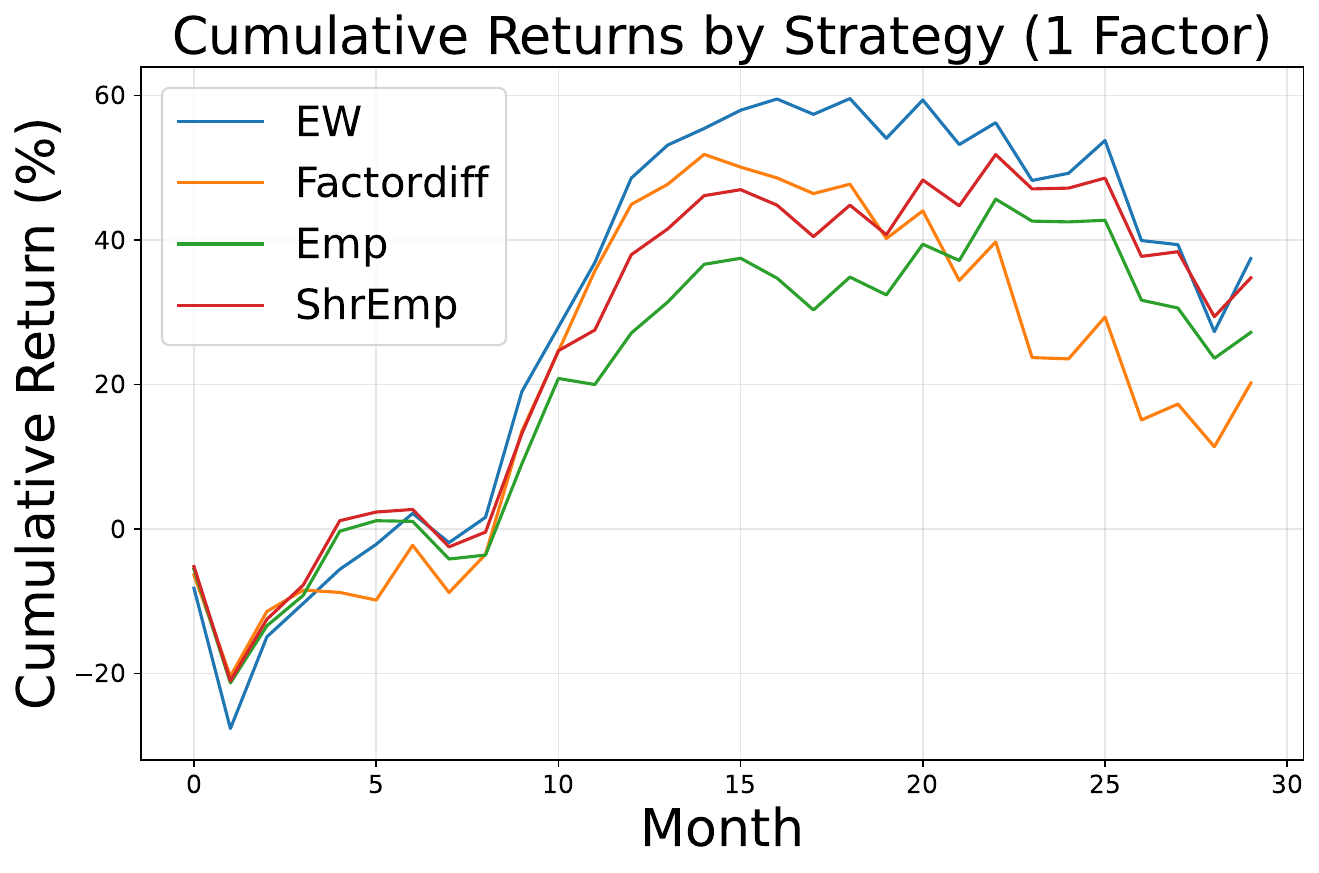}
    \caption{Low capacity (high bias)}
\end{subfigure}
\hfill
\begin{subfigure}[t]{0.32\linewidth}
    \centering
    \includegraphics[width=\linewidth]{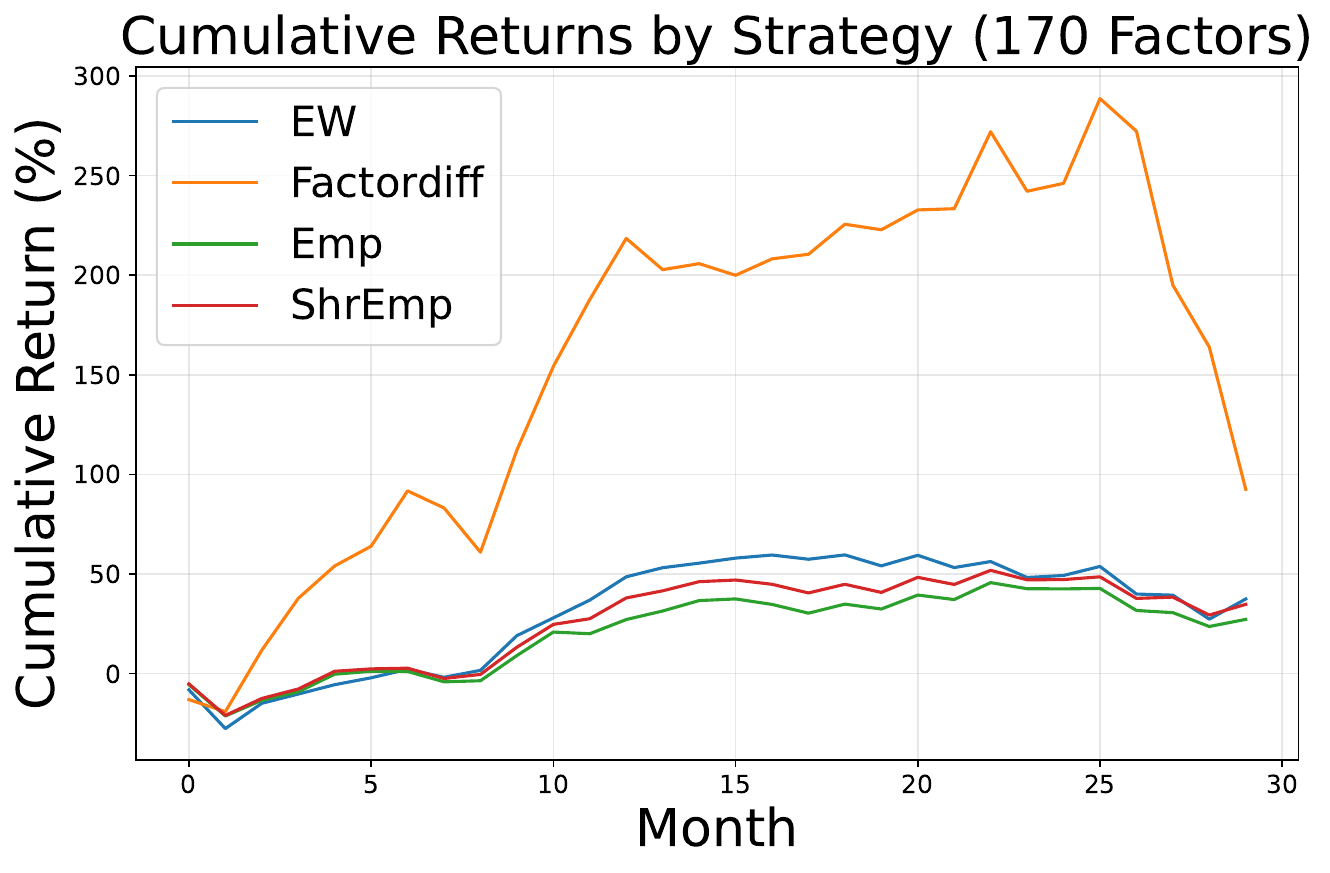}
    \caption{Medium capacity}
\end{subfigure}
\hfill
\begin{subfigure}[t]{0.32\linewidth}
    \centering
    \includegraphics[width=\linewidth]{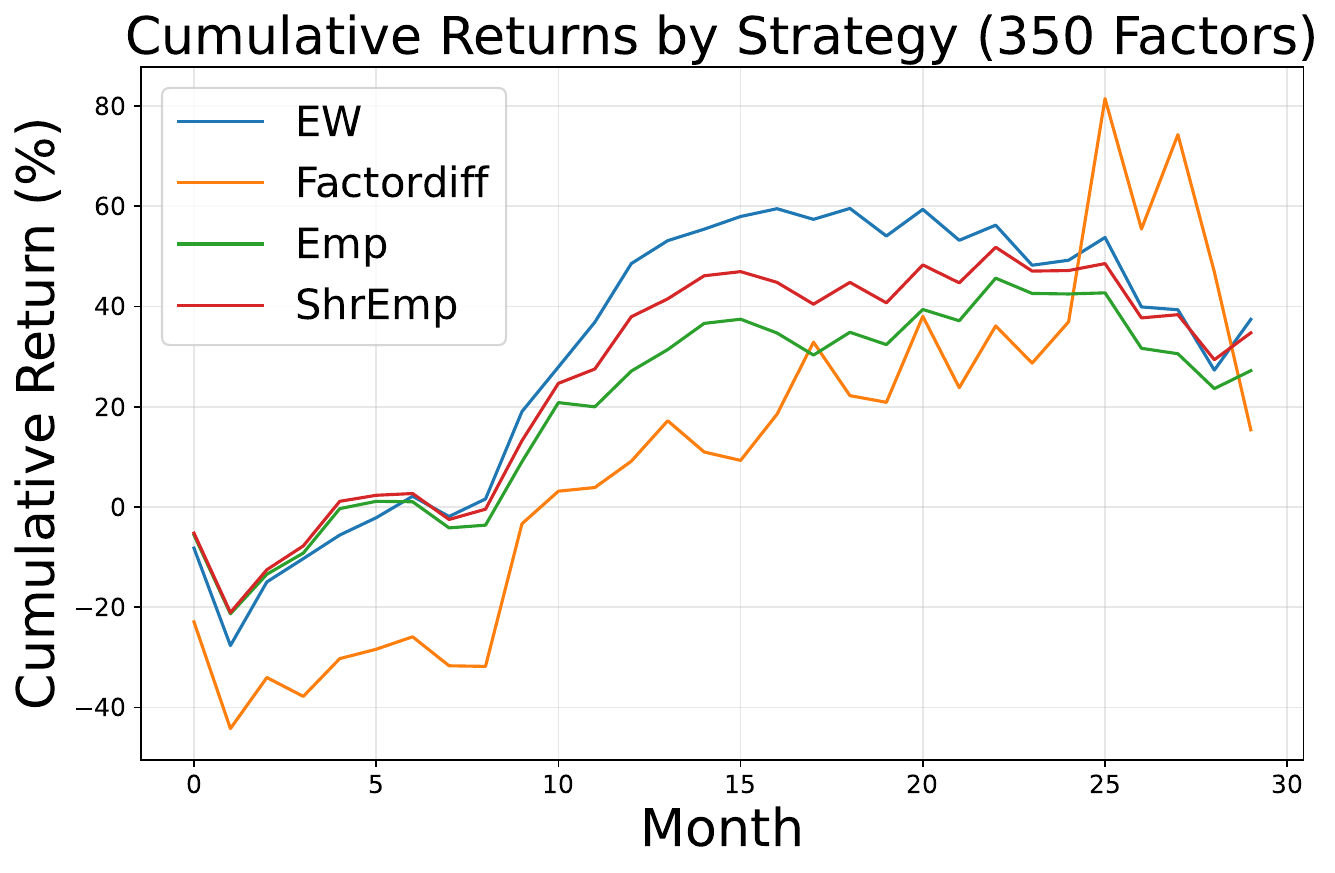}
    \caption{High capacity (high variance)}
\end{subfigure}

\caption{Cumulative portfolio return for 200 samples over the test months for four strategies. Bias--variance tradeoff illustrated through model capacity. A low-capacity model (left) underfits the data, exhibiting high bias. An intermediate model (middle) achieves a favorable
bias--variance balance and the best generalization (we verify this with a larger sample size in Figure \ref{fig:opt-long-run}). An overly expressive model (right) overfits, showing high variance
and reduced out-of-sample stability.}
\label{fig:three_panel_returns}
\end{figure}

\bibliography{iclr2026_conference}
\bibliographystyle{iclr2026_conference}

\appendix

\section{Appendix}

\subsection{Data}\label{app:data}
Our analysis uses the Global Factor Data constructed by Jensen, Kelly, Pederson and distributed through Wharton Research Data Services (WRDS). The dataset combines information from CRSP and Compustat to provide a comprehensive panel of firm-level characteristics and return for publicly traded equities. The dataset includes data from January 2010 until February 2025. The data includes more than 400 characteristics constructed following the procedures documented in \citep{JensenKellyPedersen2023}.

To align the WRDS factor data with the diffusion framework, we apply standard cross-sectional preprocessing and construct a fix-shape monthly panel. We restrict the sample to U.S. common stocks and define the prediction target as next-month return by shifting realized returns forward. Returns are winsorized cross-sectionally to mitigate outliers, while firm characteristics are imputed using cross-sectional means, standardized, and clipped within each month. For each month, we retain a fixed number of assets and organize the resulting data into tensors of characteristics and returns with dimensions $(T, N, K)$ and $(T, N)$, respectively, where $T$ denotes the number of months, $N$ denotes the number of assets, and $K$ the number of firm-level characteristics. We use $T=150$, $N = 200$, $K = 350$. These tensors serve as inputs to the conditional diffusion model.

\subsection{Related Work}
\label{app:related_work}

\paragraph{Diffusion Models:} Diffusion models learn complex data distributions by progressively corrupting data with noise and training a neural network to reverse this process. The model learns the score function via score matching, enabling sampling by iteratively denoising from noise back to data \citep{ho2020denoisingdiffusionprobabilisticmodels,song2021scorebasedgenerativemodelingstochastic}. For time series, prior work uses diffusion either (i) as a conditional scenario generator for forecasting or (ii) as a conditional model for missing-data problems \citep{rasul2021autoregressivedenoisingdiffusionmodels, tashiro2021csdiconditionalscorebaseddiffusion}. Surveys synthesize highlight evaluation pitfalls in diffusion-based time-series forecasting~\citep{meijer2024risediffusionmodelstimeseries,su2025diffusionmodelstimeseries}.

In finance, diffusion is mainly used as a conditional scenario generator for returns, with several papers emphasizing controllability or finance-specific noise structure. Shen et al. propose a non-autoregressive conditional diffusion model for generating future time-series trajectories conditioned on historical data, which can be applied to financial return forecasting \citep{shen2023nonautoregressiveconditionaldiffusionmodels}. Tanaka et al.~focus on controllable conditional generation for financial time series, adding explicit controls to steer the sampled trajectories toward desired attributes~\cite{tanaka2025cofindiffcontrollablefinancialdiffusion}. Kim et al.\ modify the diffusion forward noising process to reflect  financial structure (e.g., heteroskedasticity and multiplicative noise), targeting more realistic synthetic dynamics and improved conditional sampling~\cite{kim2025diffusionbasedgenerativemodelfinancial}. Wang et al.\ study finance-tailored denoisers and Takahashi et al.\ propose methods aimed at synthetic financial time-series generation with finance-specific modeling choices~\citep{wang2024financialtimeseriesdenoiser,takahashi2024generationsyntheticfinancialtime}. Beyond return/path generation, Jin et al.\ apply diffusion in an option-centric setting by forecasting the implied-volatility surface~\citep{jin2025forecastingimpliedvolatilitysurface}.

\paragraph{Factor Models}
Factor models are a standard framework for portfolio construction, with widely used specifications such \citet{fama1993common,fama2015five,carhart1997persistence}. Modern portfolio risk systems build on this framework by estimating factor exposures and covariance structures \citet{rosenberg1974extra} and its practical development for quantitative portfolio construction \citet{grinold2000barra}. A limitation of factor models is estimation error in high dimensions \citep{ledoit2004honey,fan2008high,fan2013largecovarianceestimationthresholding}. Work in empirical asset pricing shows that large panels of firm characteristics improve return prediction but introduce redundancy and model selection challenges \citep{gu2020empirical,kelly2019ipca}. As the number of proposed factors has expanded, recent studies emphasize systematic testing, dimensionality control, and the risk of overfitting in high-dimensional factor spaces \citep{feng2026selectingtestingassetpricing,borri2025forwardselectionfamamacbethregression}.

\subsection{Portfolio Constructions}\label{app:portfolio_constructions}

In the transaction cost setting, we augment the objective with linear trading costs. Portfolio returns are computed as the inner product of portfolio weights and realized returns, and performance is summarized using mean return, volatility, and annualized Sharpe ratio.

\begin{itemize}
    \item \textbf{Equal-Weighted (EW)} assigns uniform weights and does not estimate return moments.
    \item \textbf{Empirical (Emp)} estimates mean and covariance directly from historical returns using a rolling window.
    \item \textbf{Shrinkage Empirical (ShrEmp)} applies covariances shrinkage to improve stability while retaining the sample covariance mean \citep{JamesStein1992}.
\end{itemize}

We obtain moment estimates the conditional distribution of next-period returns using the conditional diffusion model outlined above. Monte Carlo samples drawn from this distribution are used to estimate the conditional mean and covariance of returns. We use 200 samples in the results diagrams.

\section{Additional Figures}\label{app:additionalgraphs}

Let $k$ denote the number of factors. We perform an ablation for $k \in \{1, 3, 6, 11, 18, 30, 48, 75, 115, 170, 240, 300, 350\}$. 

\subsection{Cumulative Returns}\label{app:cumm_ret}

The figures in this section illustrate how performance varies as the number of factors
$k$ increases. When $k$ is small, the model is overly constrained and fails to capture sufficient structure in the data. This high-bias regime leads to underfitting: cumulative returns closely track or underperform the baseline. As $k$ increases, performance improves and the model begins to capture meaningful relationships. It begins to out-perform the baseline when $k \geq 18$. However, for very large $k$, performance deteriorates again. The model enters a high-variance regime in which additional factors primarily fit noise rather than signal. This is reflected in reduced out-of-sample performance, with returns again failing to outperform the baseline.

\begin{figure}[H]
\centering
\includegraphics[width=0.9\linewidth]{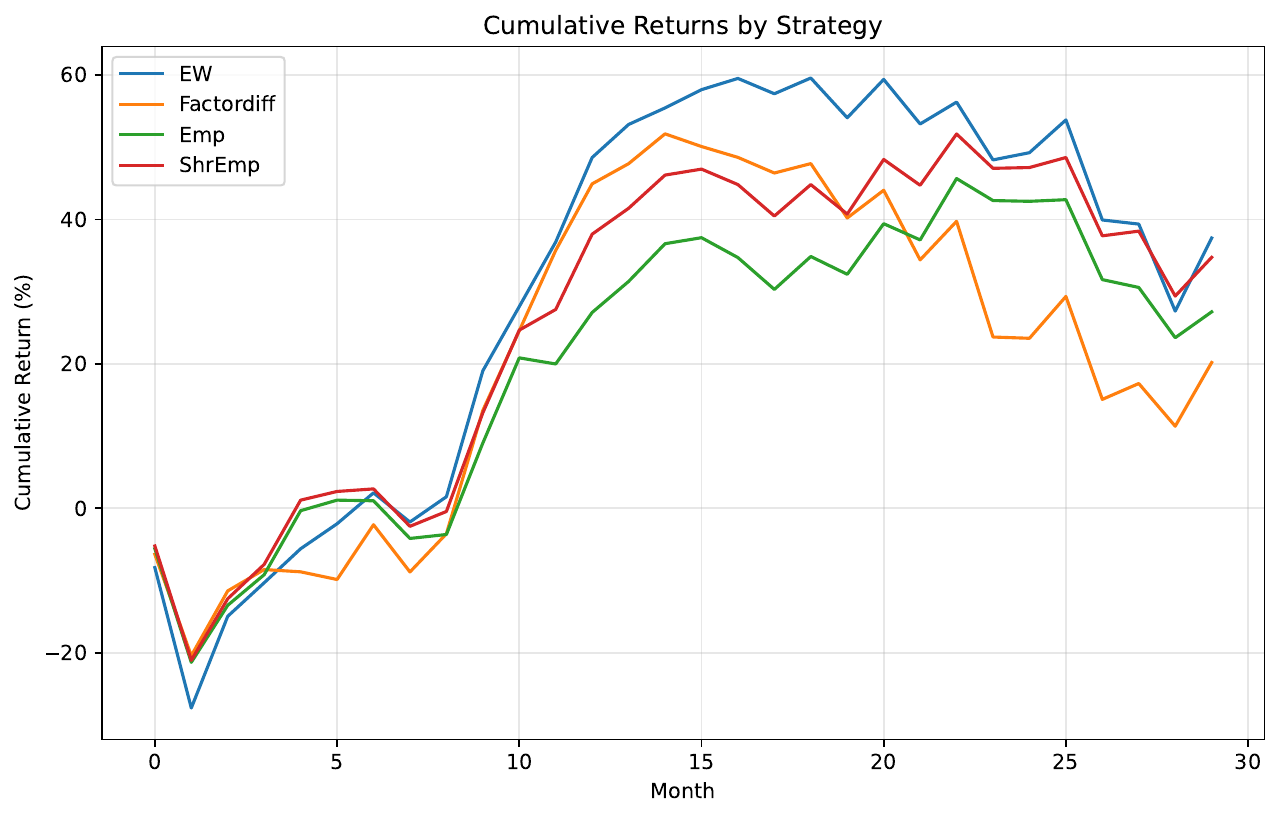}
\caption{Cumulative returns ($k=1$)}
\end{figure}

\begin{figure}[H]
\centering
\includegraphics[width=0.9\linewidth]{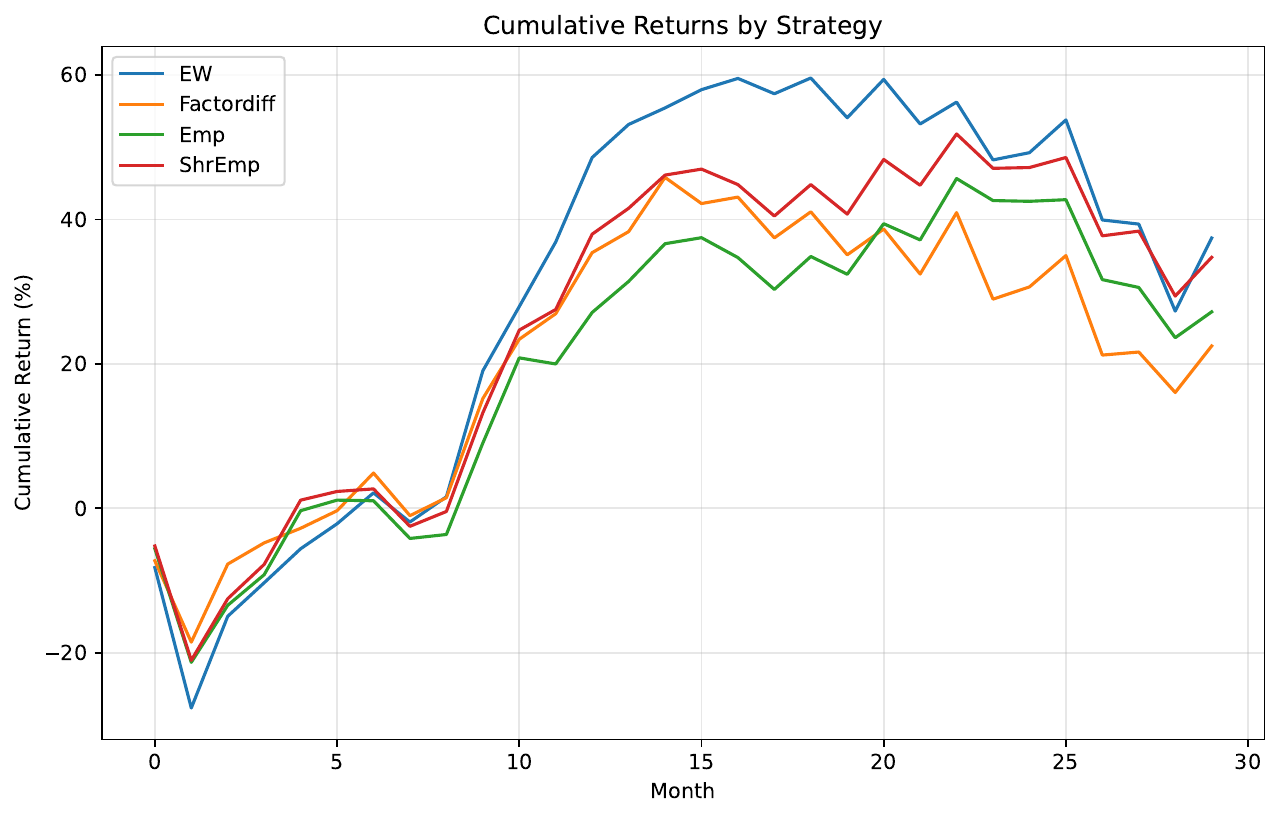}
\caption{Cumulative returns ($k=3$)}
\end{figure}

\begin{figure}[H]
\centering
\includegraphics[width=0.9\linewidth]{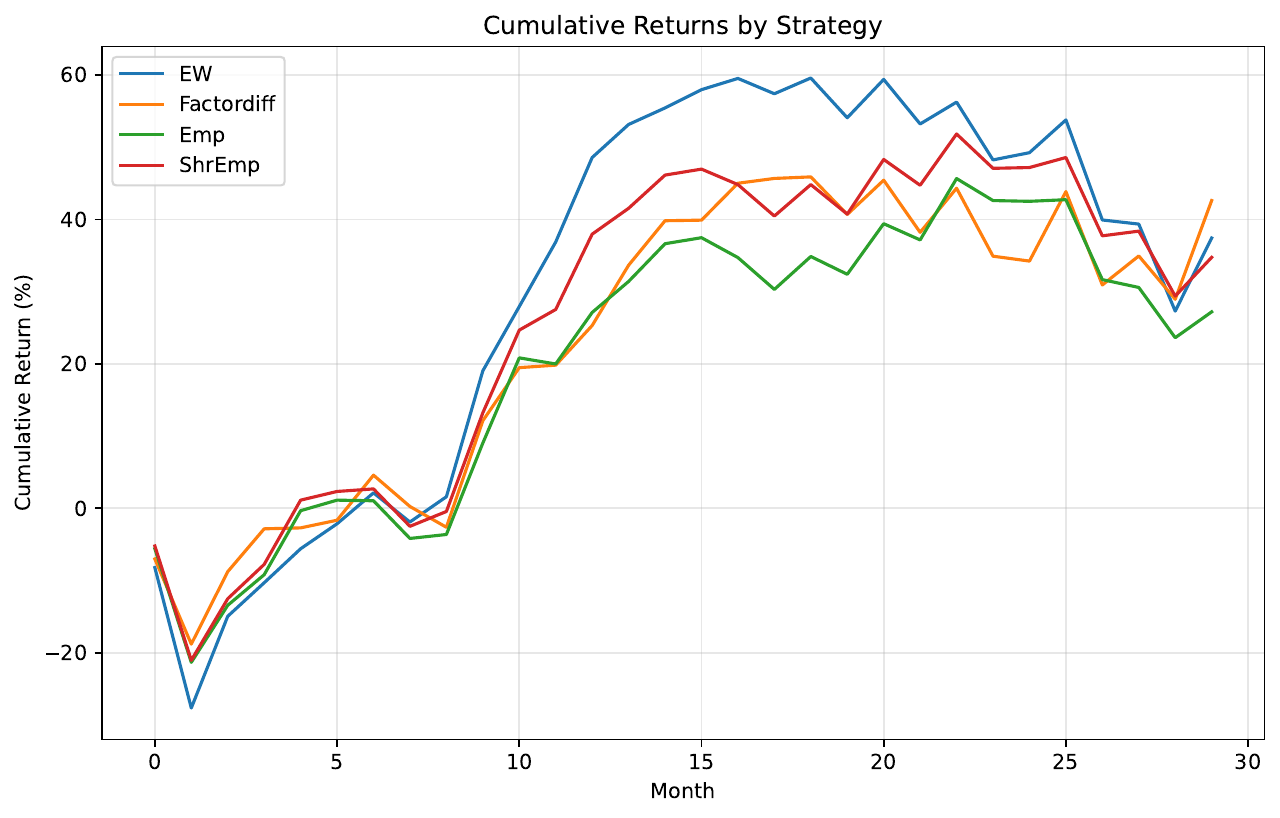}
\caption{Cumulative returns ($k=6$)}
\end{figure}

\begin{figure}[H]
\centering
\includegraphics[width=0.9\linewidth]{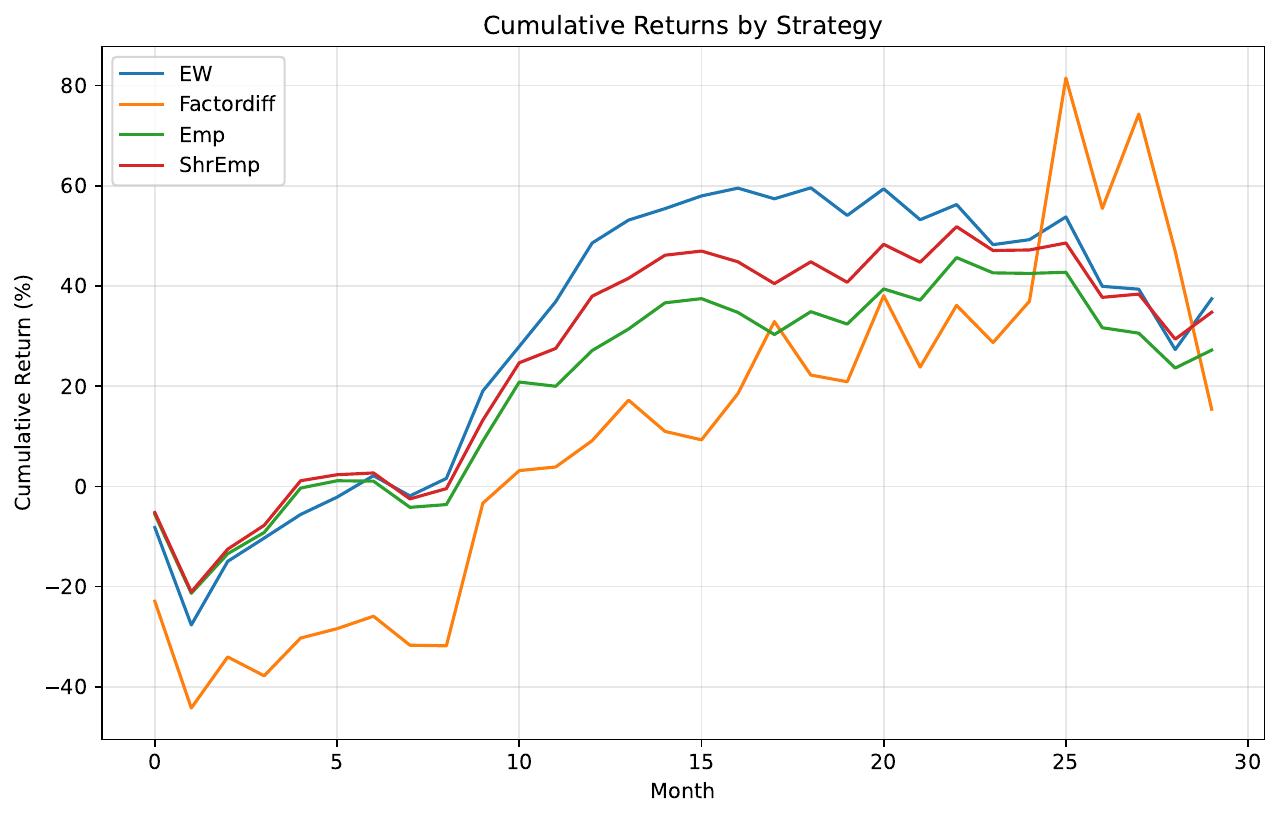}
\caption{Cumulative returns ($k=10$)}
\end{figure}

\begin{figure}[H]
\centering
\includegraphics[width=0.9\linewidth]{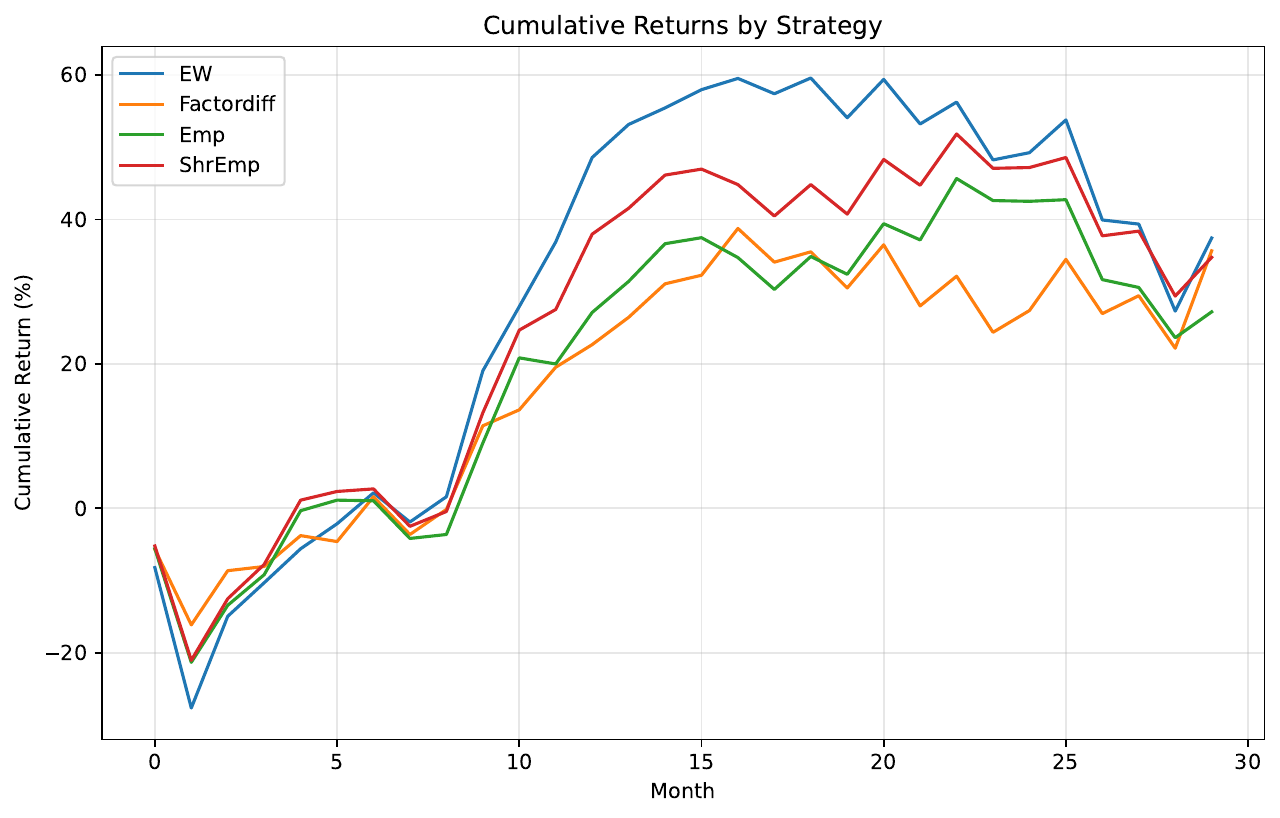}
\caption{Cumulative returns ($k=11$)}
\end{figure}

\begin{figure}[H]
\centering
\includegraphics[width=0.9\linewidth]{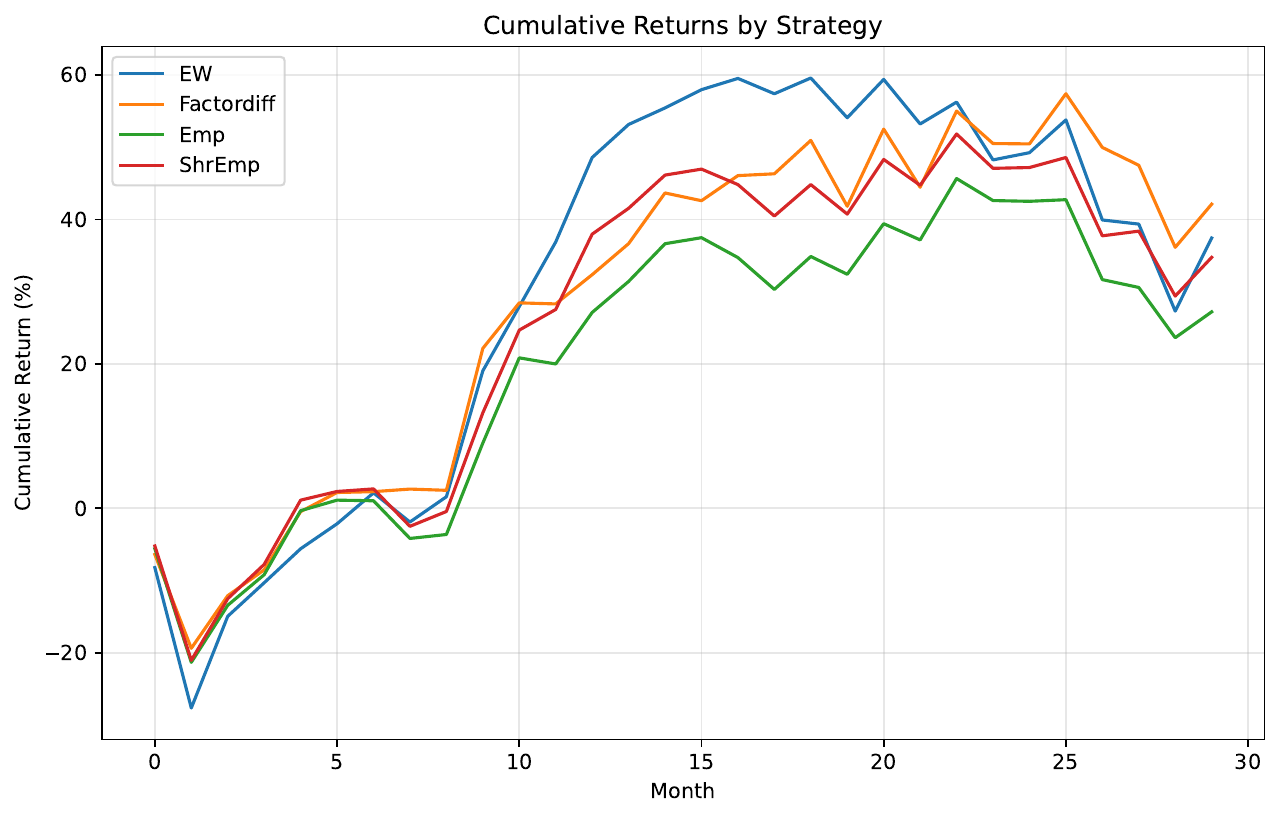}
\caption{Cumulative returns ($k=18$)}
\end{figure}

\begin{figure}[H]
\centering
\includegraphics[width=0.9\linewidth]{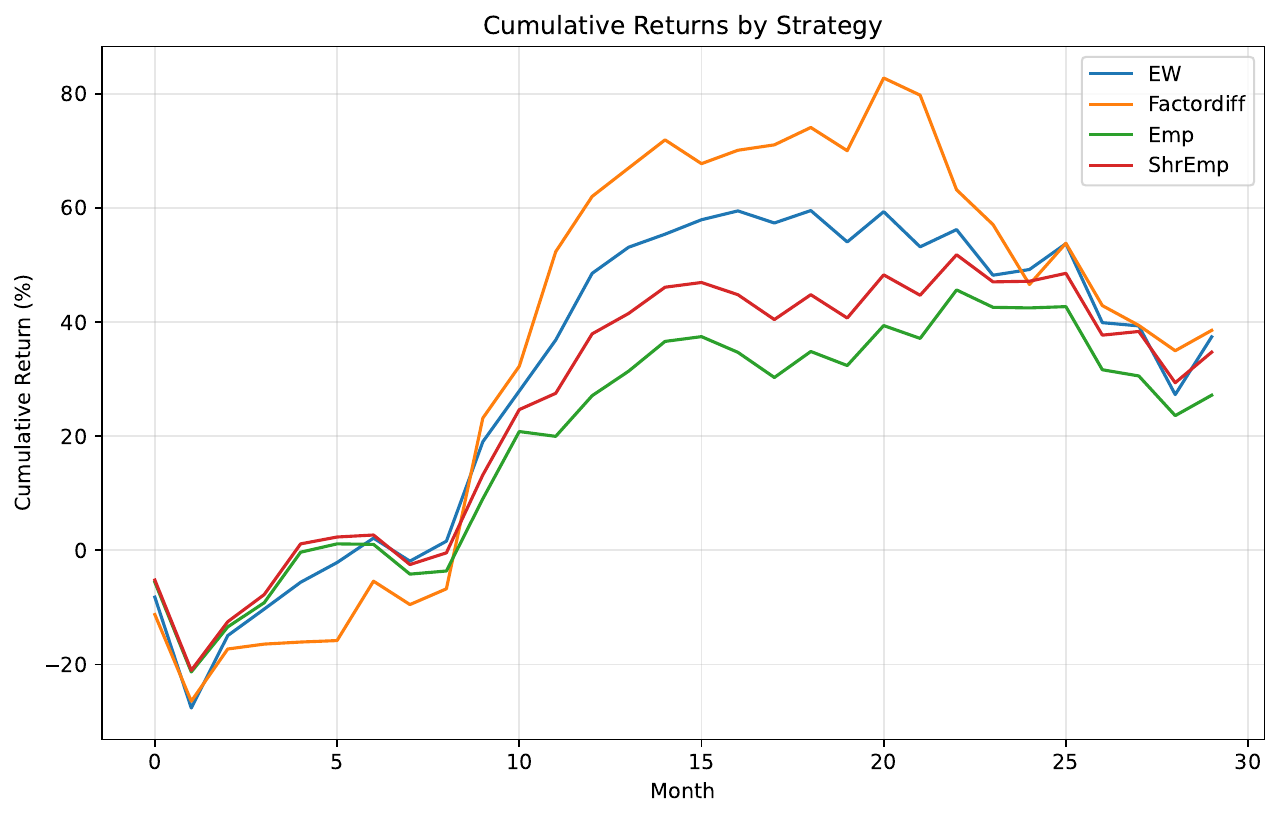}
\caption{Cumulative returns ($k=30$)}
\end{figure}

\begin{figure}[H]
\centering
\includegraphics[width=0.9\linewidth]{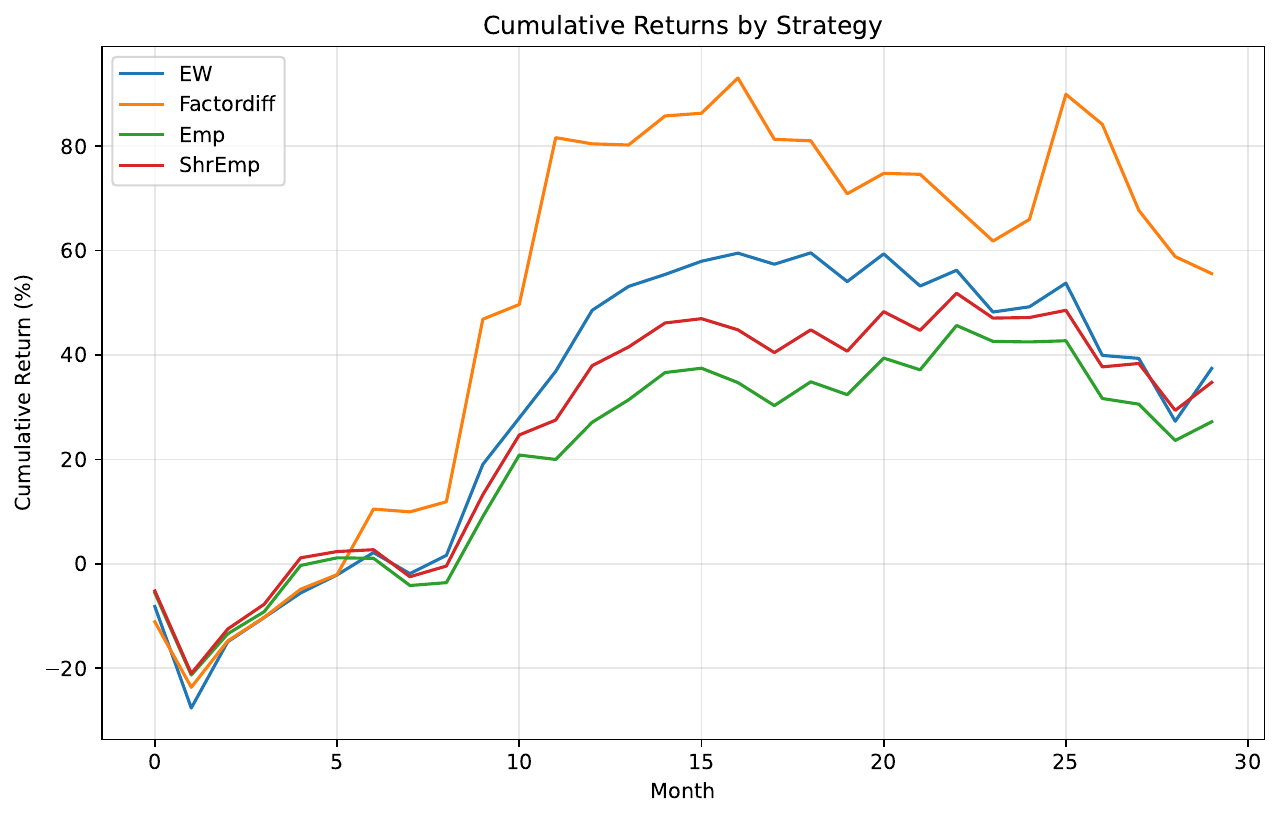}
\caption{Cumulative returns ($k=48$)}
\end{figure}

\begin{figure}[H]
\centering
\includegraphics[width=0.9\linewidth]{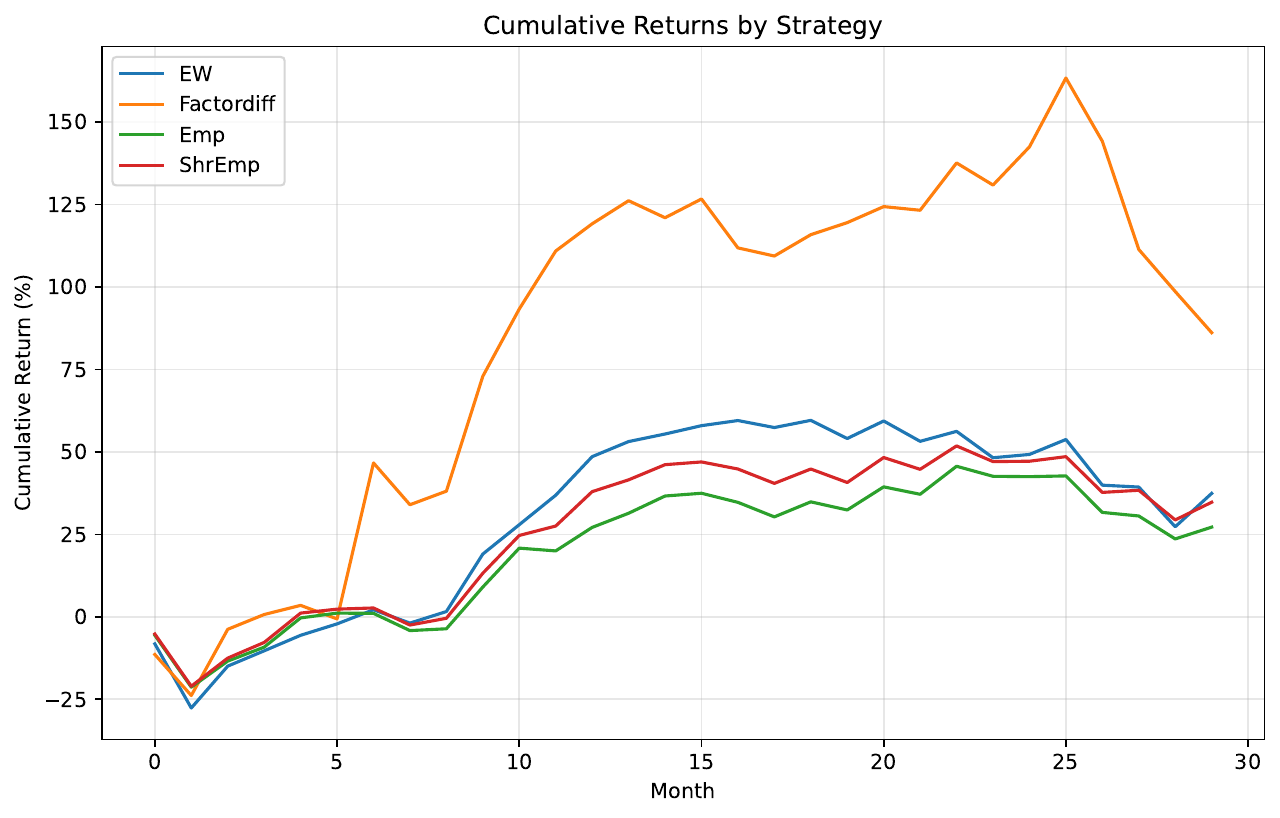}
\caption{Cumulative returns ($k=75$)}
\end{figure}

\begin{figure}[H]
\centering
\includegraphics[width=0.9\linewidth]{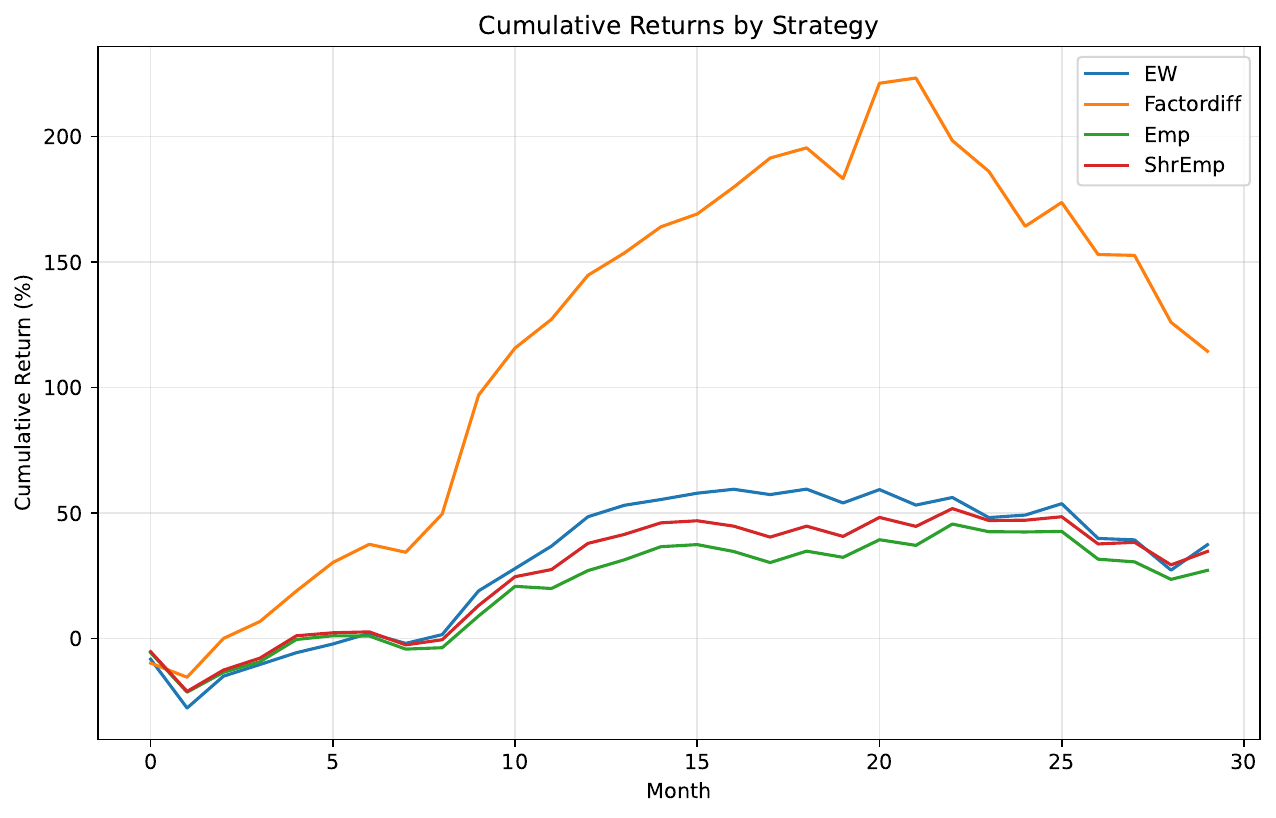}
\caption{Cumulative returns ($k=115$)}
\end{figure}

\begin{figure}[H]
\centering
\includegraphics[width=0.9\linewidth]{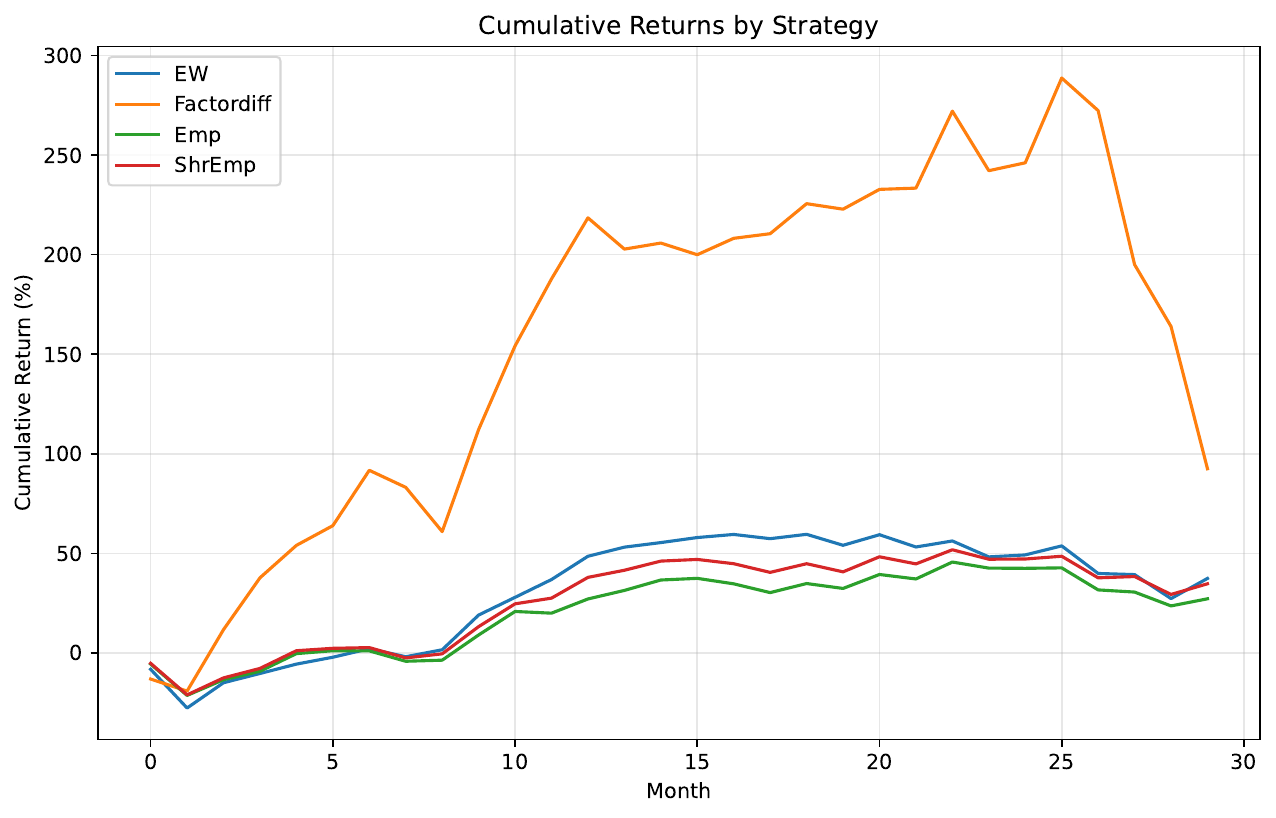}
\caption{Cumulative returns ($k=170$)}
\end{figure}

\begin{figure}[H]
\centering
\includegraphics[width=0.9\linewidth]{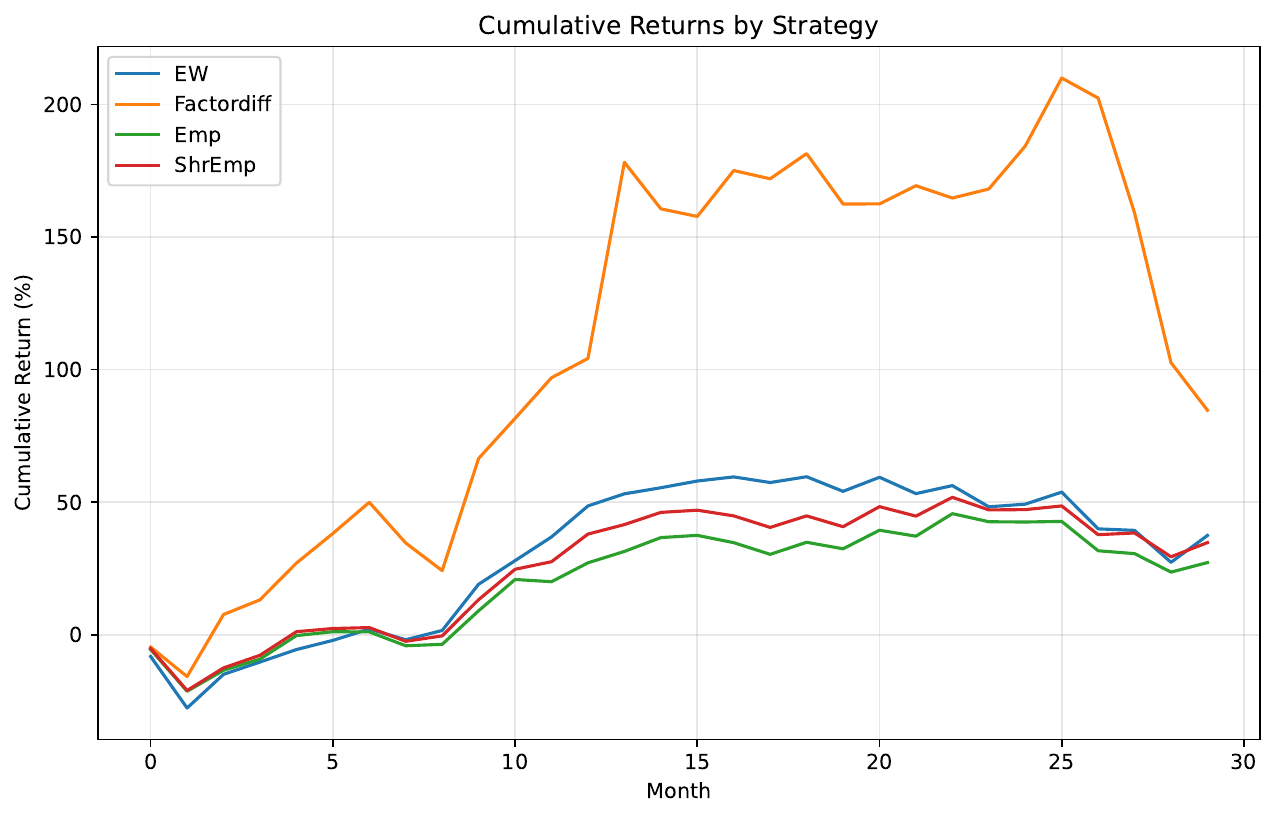}
\caption{Cumulative returns ($k=240$)}
\end{figure}

\begin{figure}[H]
\centering
\includegraphics[width=0.9\linewidth]{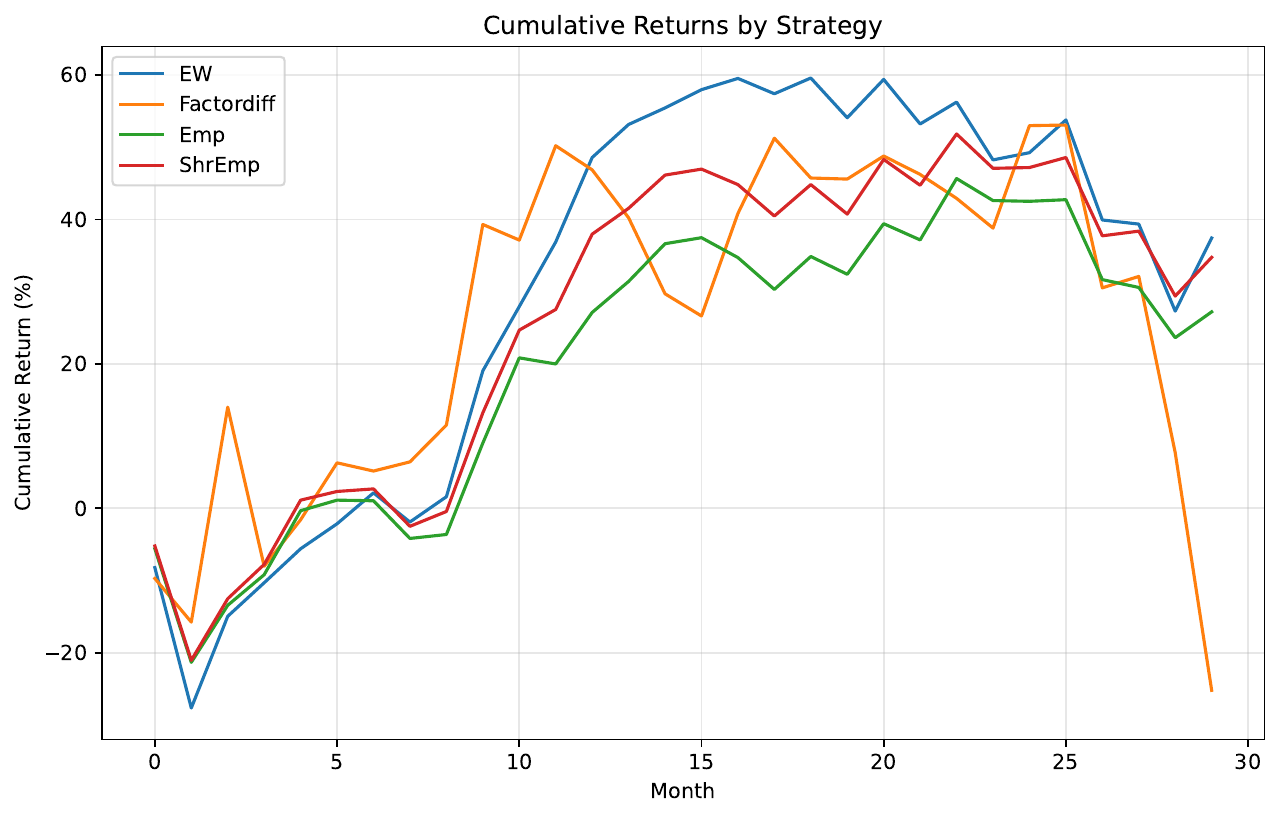}
\caption{Cumulative returns ($k=300$)}
\end{figure}

\begin{figure}[H]
\centering
\includegraphics[width=0.9\linewidth]{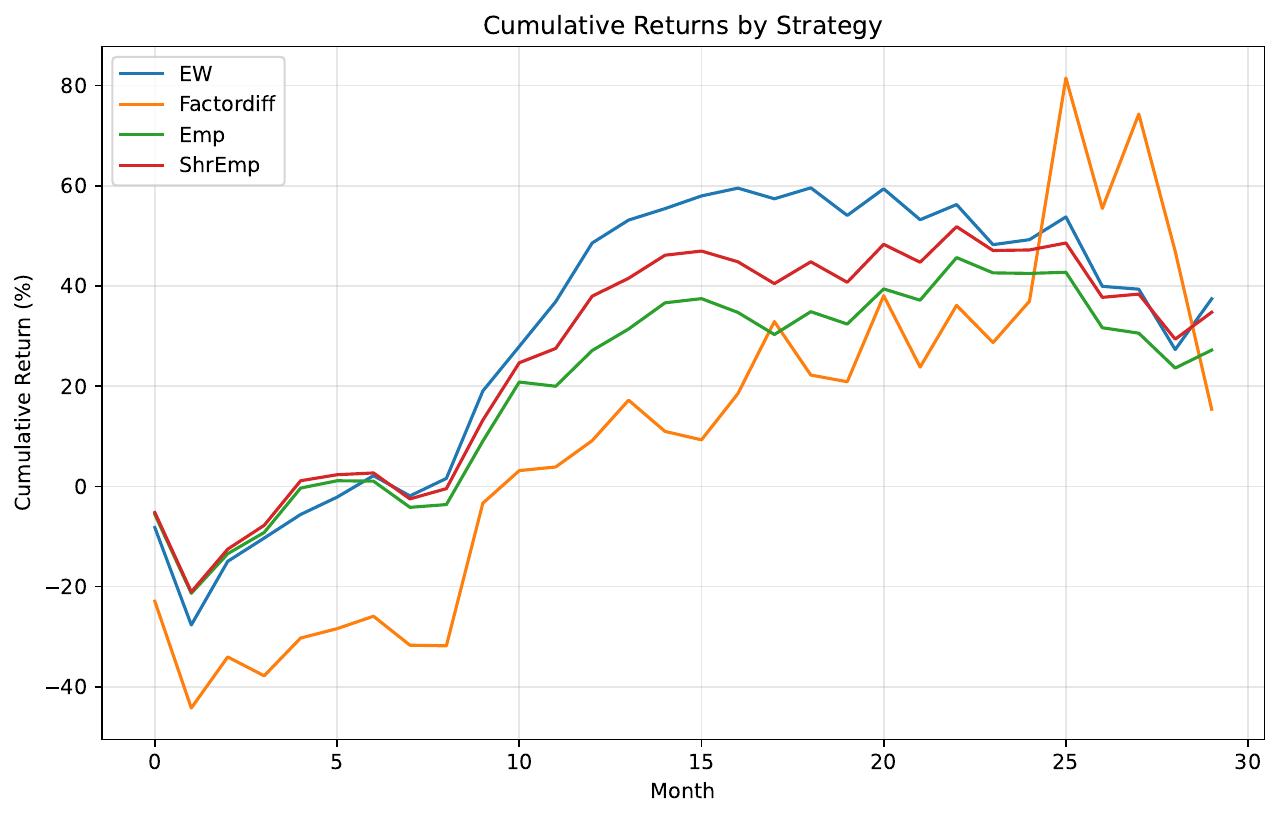}
\caption{Cumulative returns ($k=350$)}
\end{figure}

We verify the performance of $k=170$, raising the number of samples to 1000, again observing that it outperforms the baseline.

\begin{figure}[H]
\centering
\includegraphics[width=0.9\linewidth]{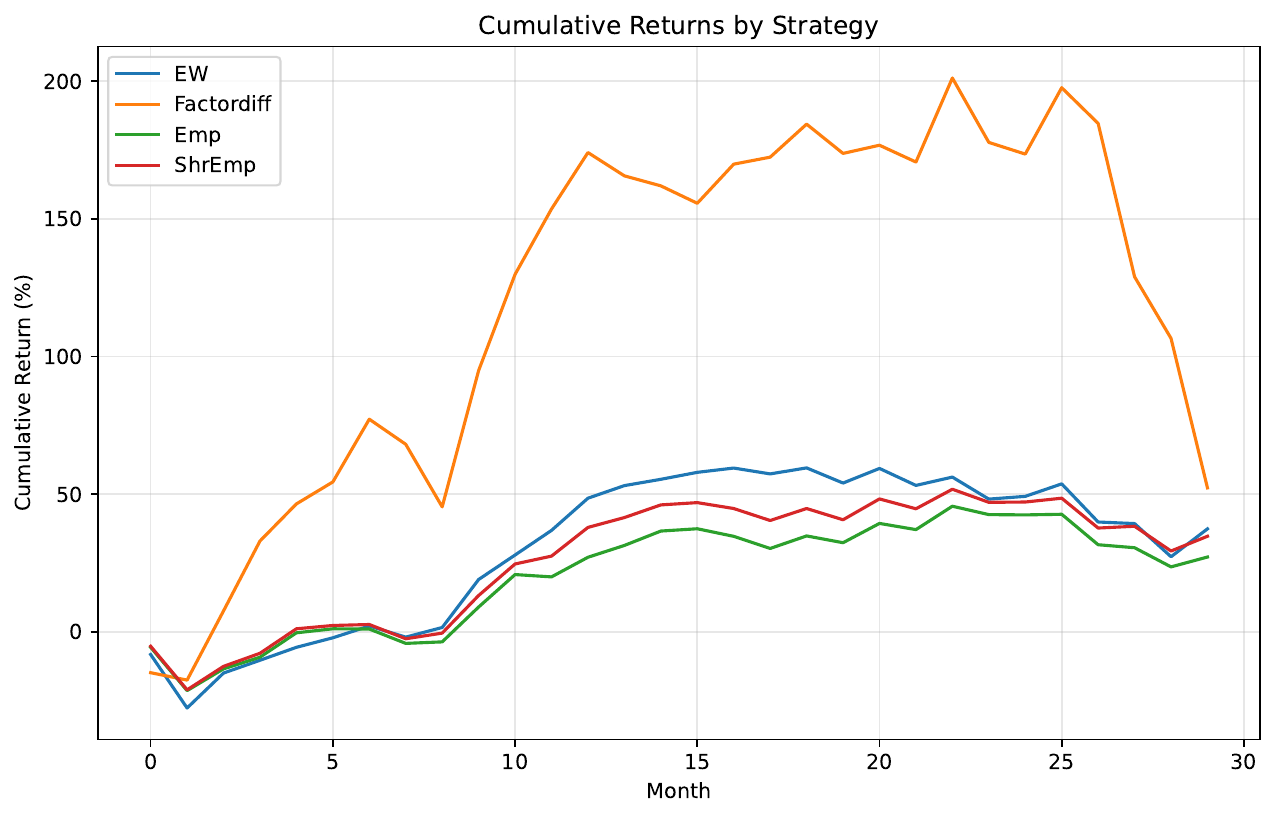}
\caption{Cumulative returns ($k=170$)}
\label{fig:opt-long-run}
\end{figure}

\clearpage
\subsection{Portfolio Weights}\label{app:port_weight}

In this section, we show how increasing the number of factors $k$ progressively changes the structure of the learned portfolios. For small $k$, the weights are broadly distributed across many assets,
indicating a diffuse allocation consistent with an under-parameterized model. As $k$ grows, the portfolio weights become increasingly concentrated, with larger magnitudes assigned to a smaller subset of assets. This concentration suggests that higher-capacity models identify more specific signals but also become more sensitive to noise, leading to over-specialized allocations and reduced out-of-sample performance.

\begin{figure}[H]
\centering
\includegraphics[width=0.9\linewidth]{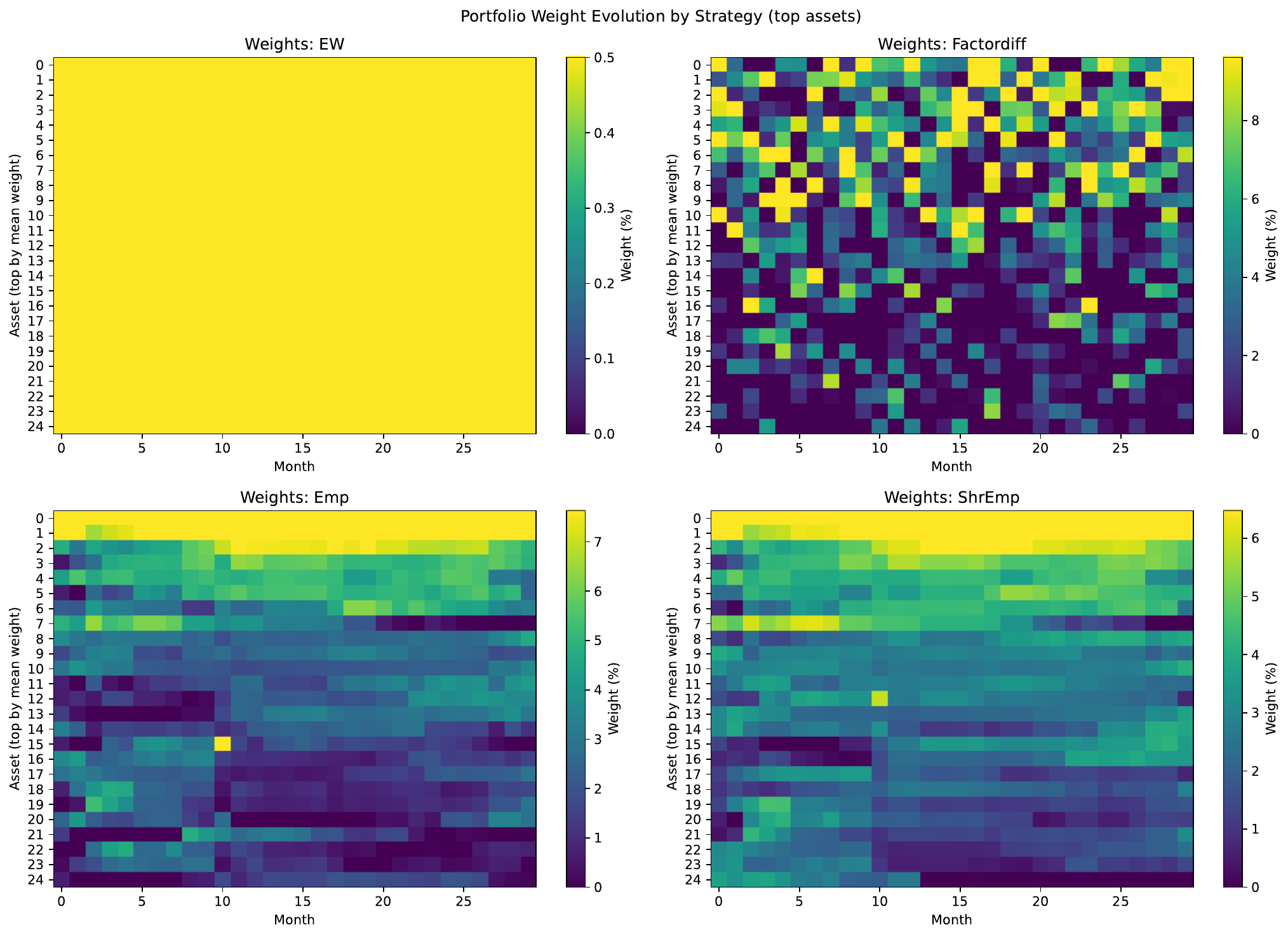}
\caption{Weights heatmap ($k=1$)}
\end{figure}

\begin{figure}[H]
\centering
\includegraphics[width=0.9\linewidth]{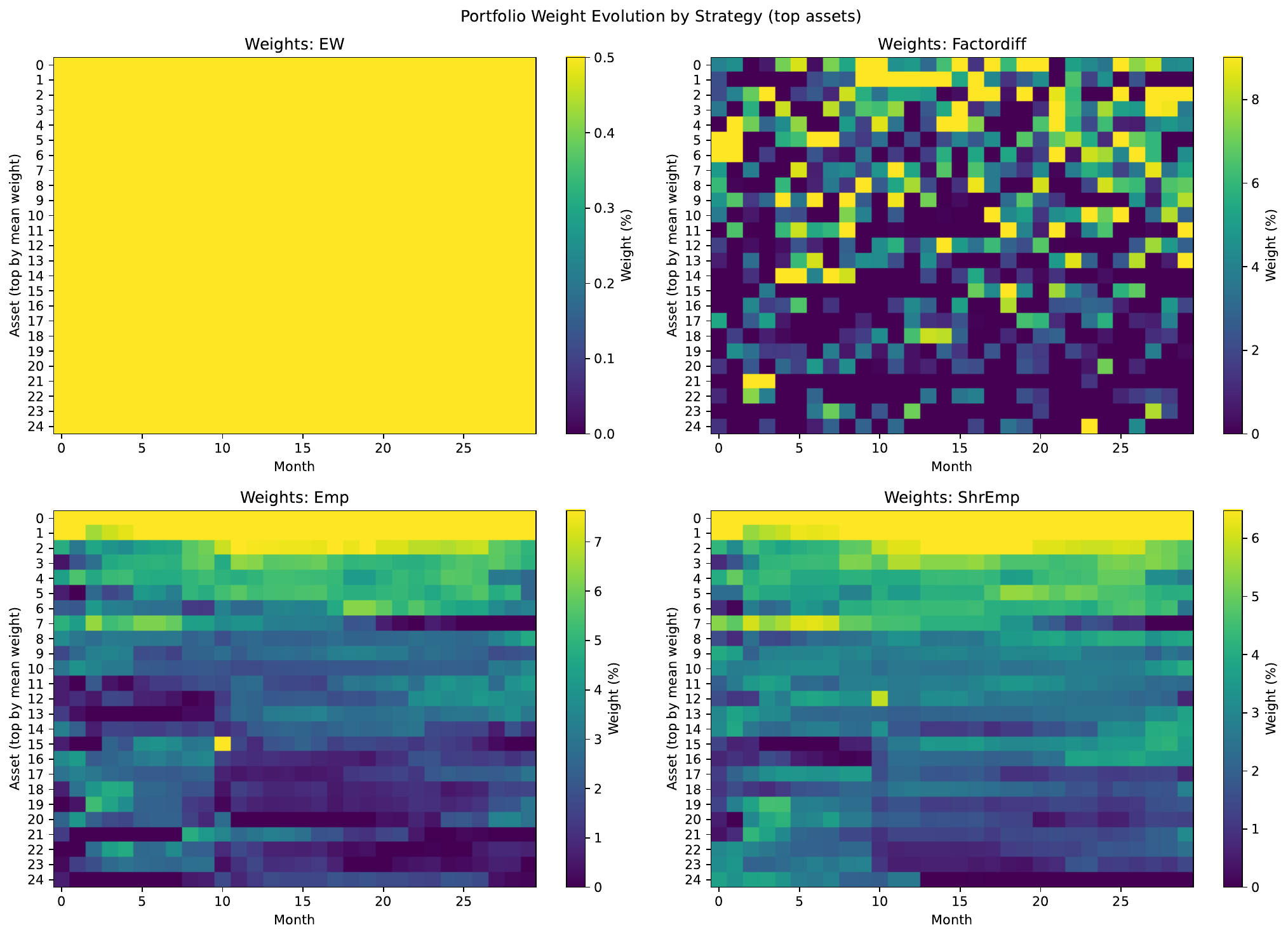}
\caption{Weights heatmap ($k=3$)}
\end{figure}

\begin{figure}[H]
\centering
\includegraphics[width=0.9\linewidth]{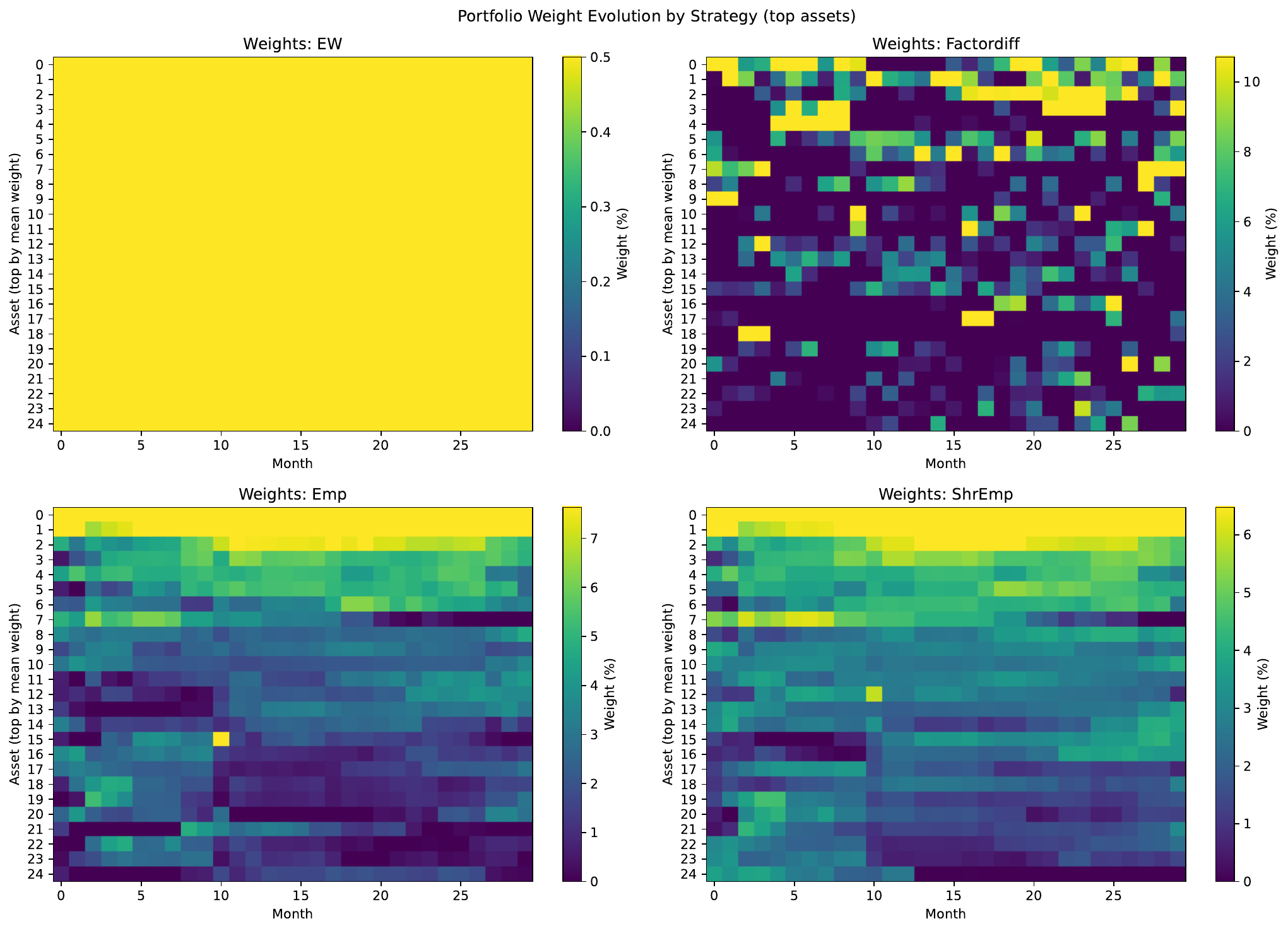}
\caption{Weights heatmap ($k=6$)}
\end{figure}

\begin{figure}[H]
\centering
\includegraphics[width=0.9\linewidth]{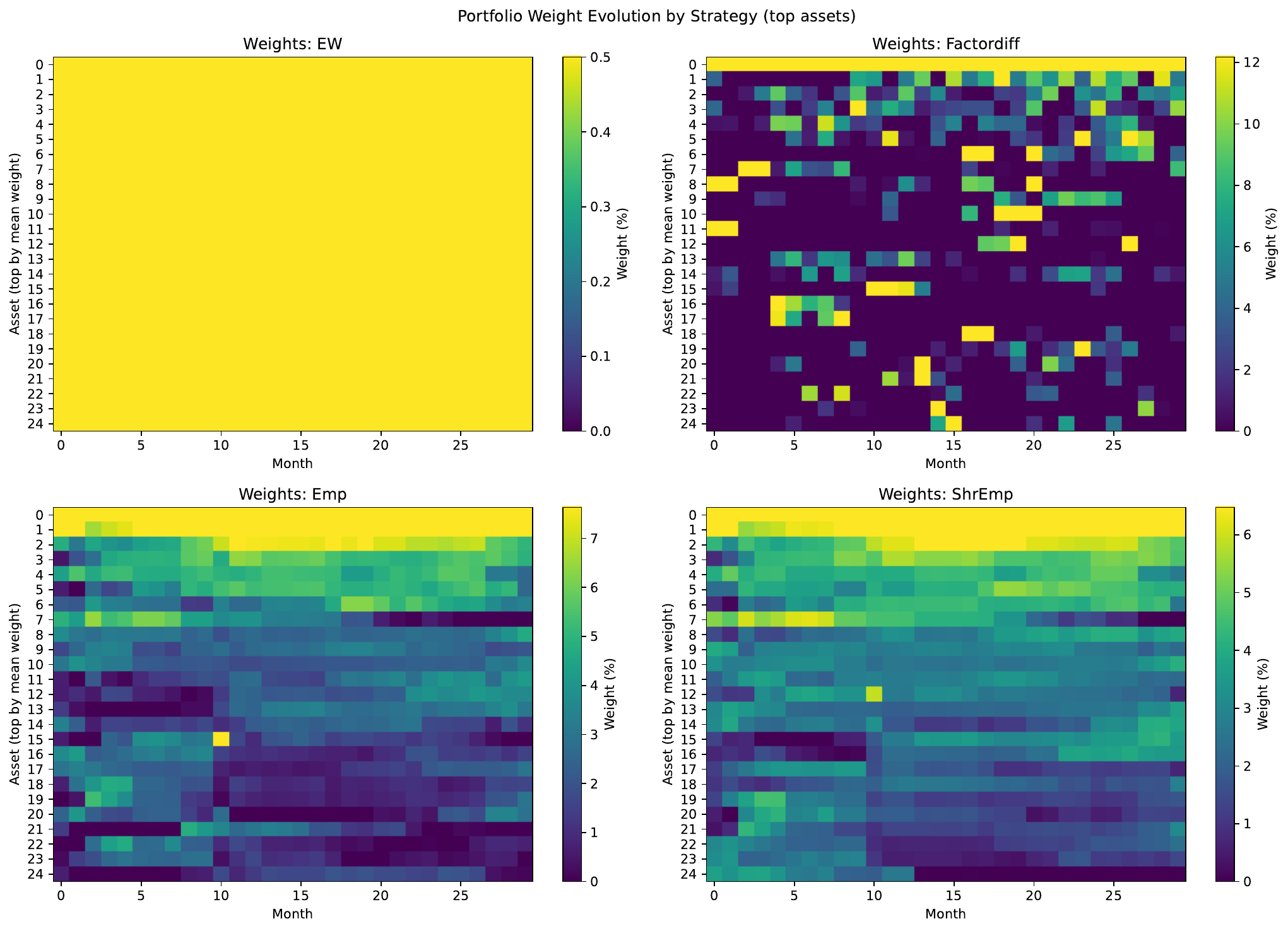}
\caption{Weights heatmap ($k=11$)}
\end{figure}

\begin{figure}[H]
\centering
\includegraphics[width=0.9\linewidth]{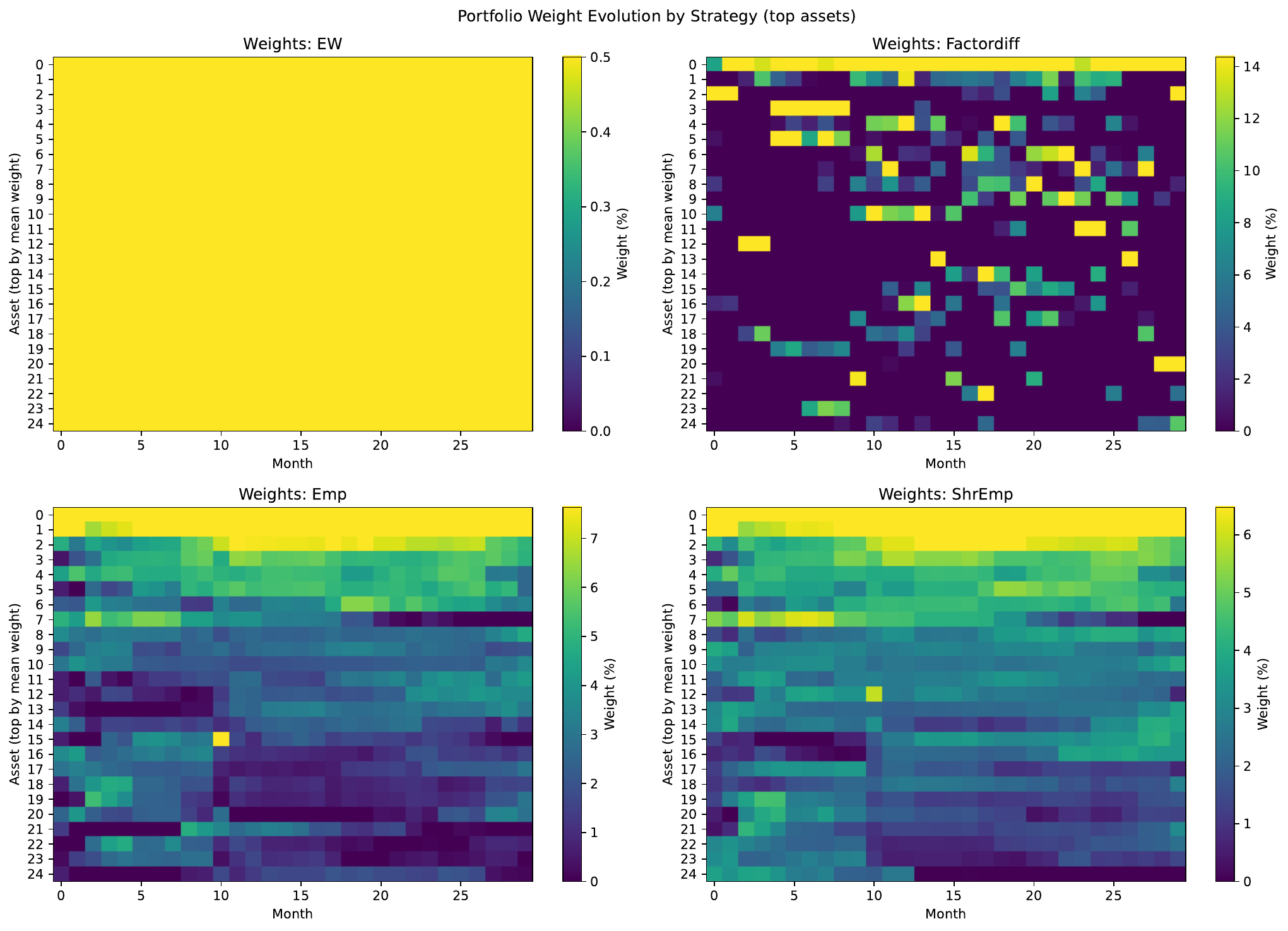}
\caption{Weights heatmap ($k=18$)}
\end{figure}

\begin{figure}[H]
\centering
\includegraphics[width=0.9\linewidth]{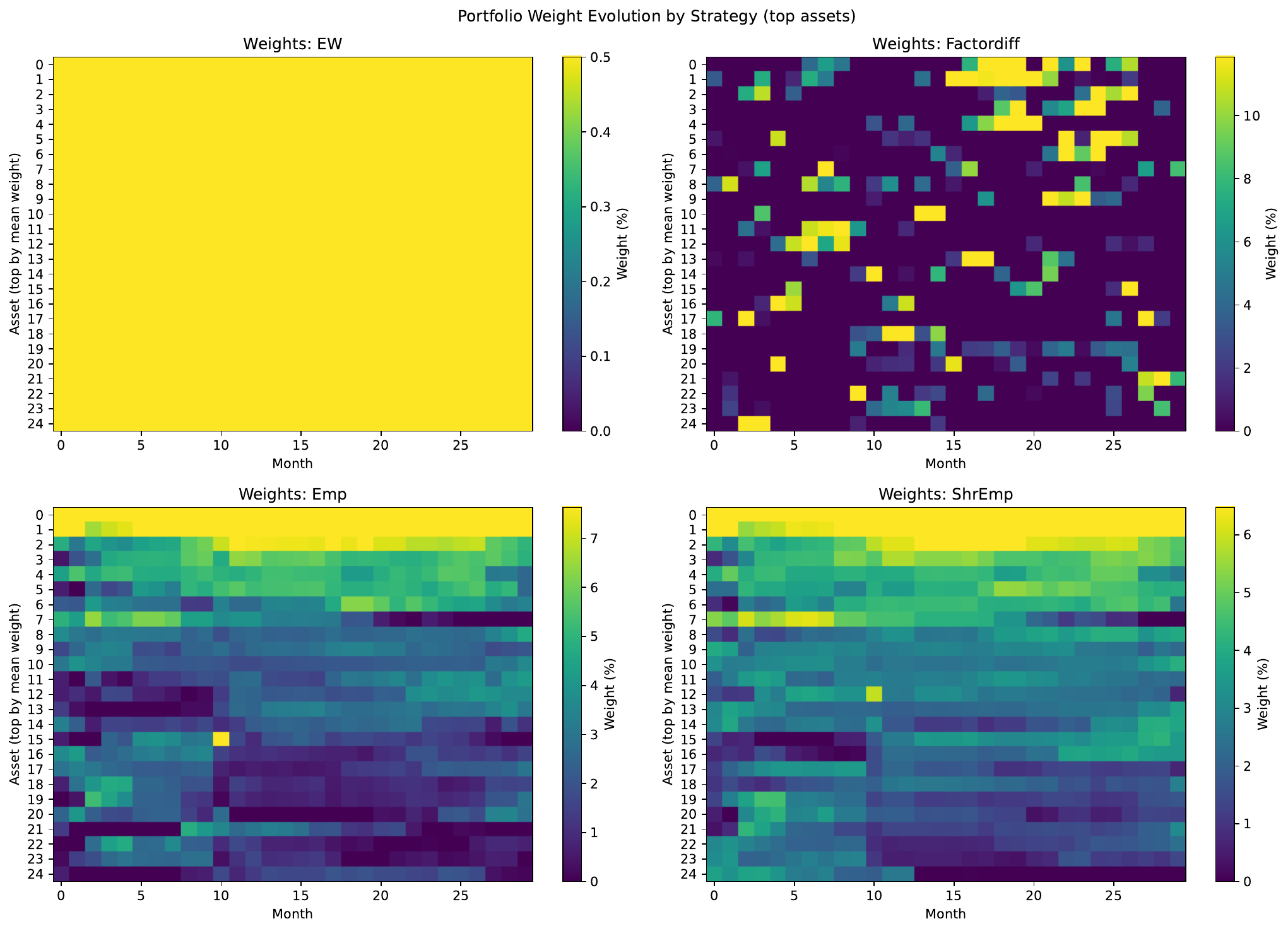}
\caption{Weights heatmap ($k=30$)}
\end{figure}

\begin{figure}[H]
\centering
\includegraphics[width=0.9\linewidth]{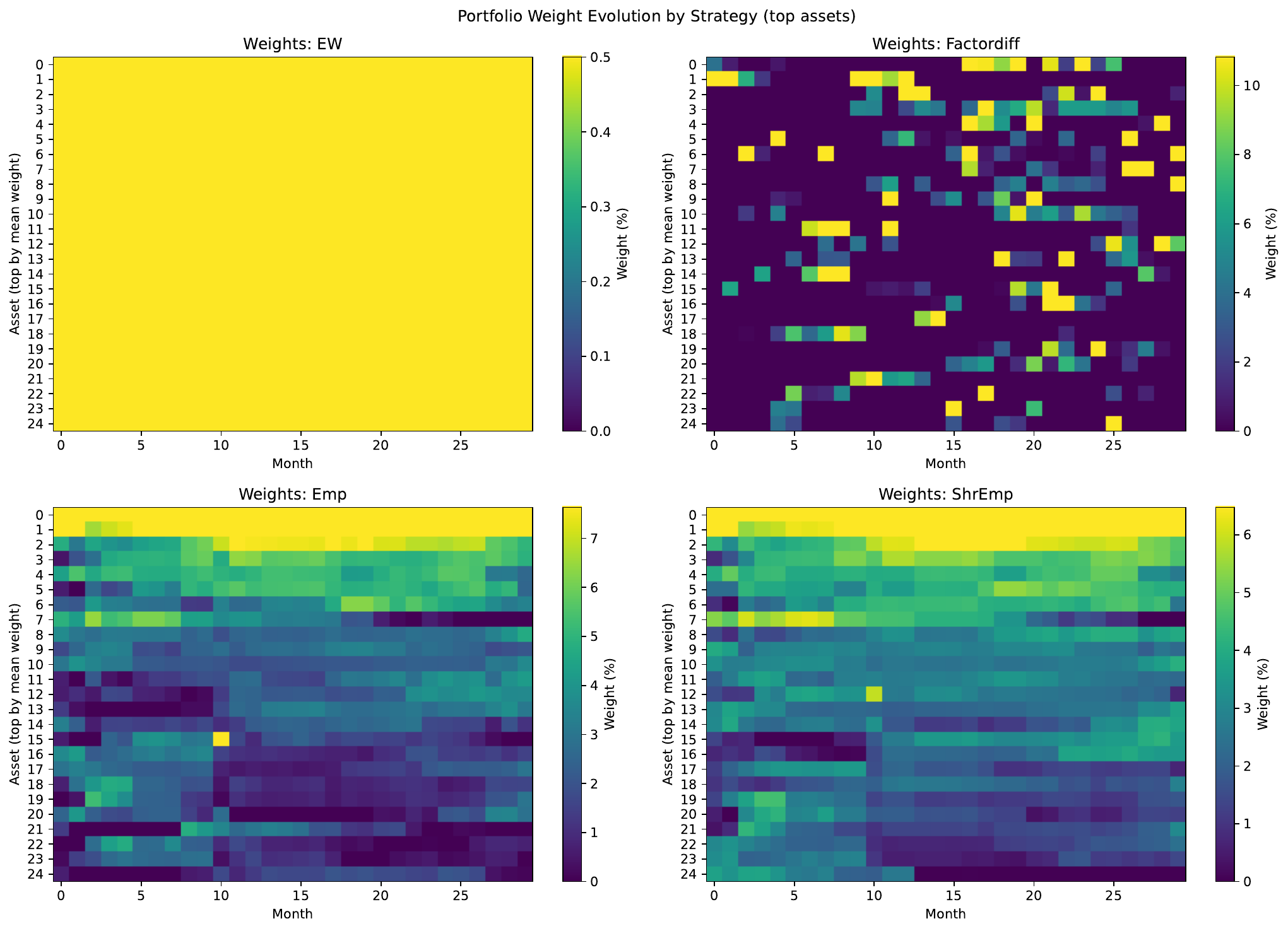}
\caption{Weights heatmap ($k=48$)}
\end{figure}

\begin{figure}[H]
\centering
\includegraphics[width=0.9\linewidth]{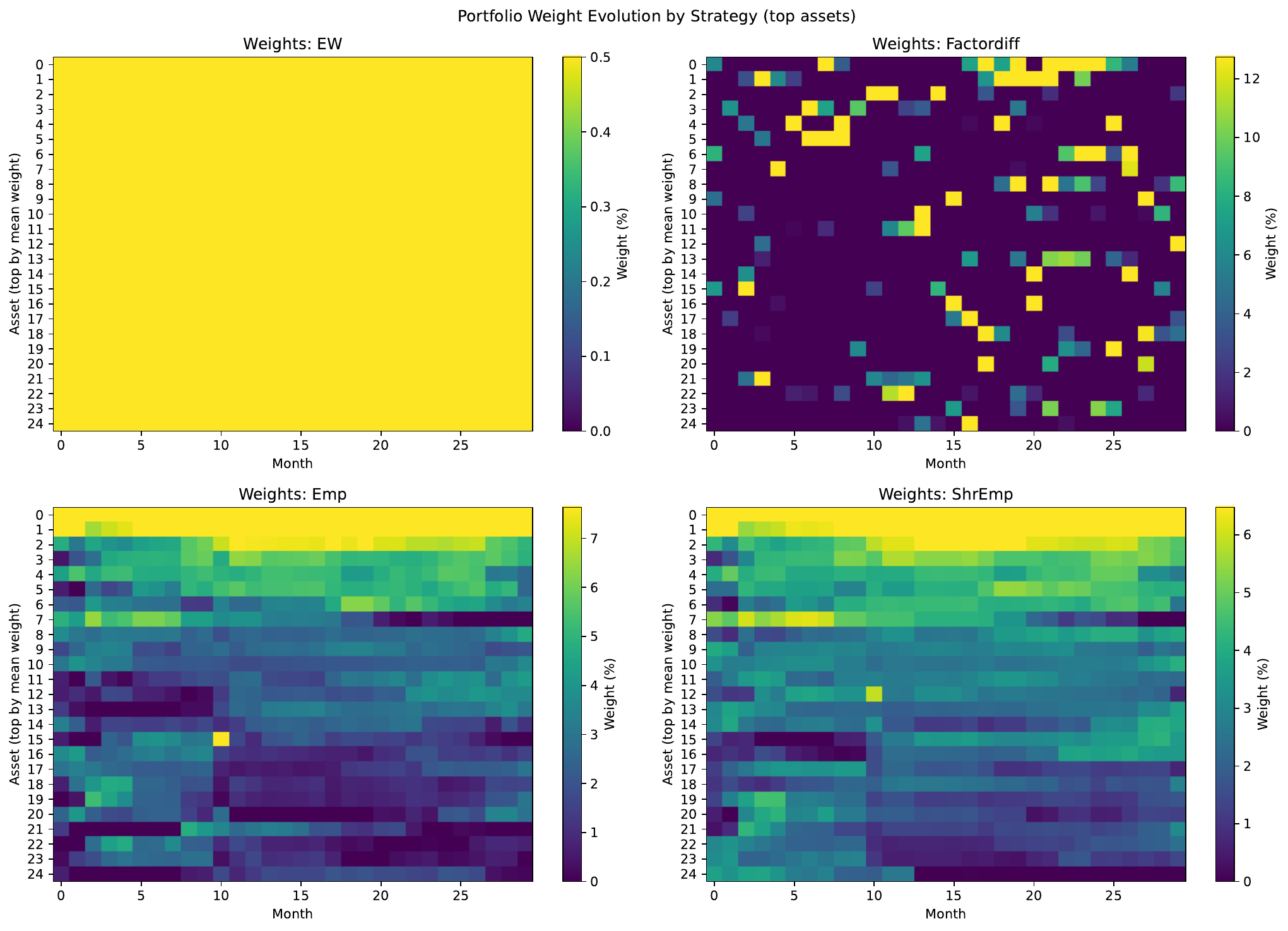}
\caption{Weights heatmap ($k=75$)}
\end{figure}

\begin{figure}[H]
\centering
\includegraphics[width=0.9\linewidth]{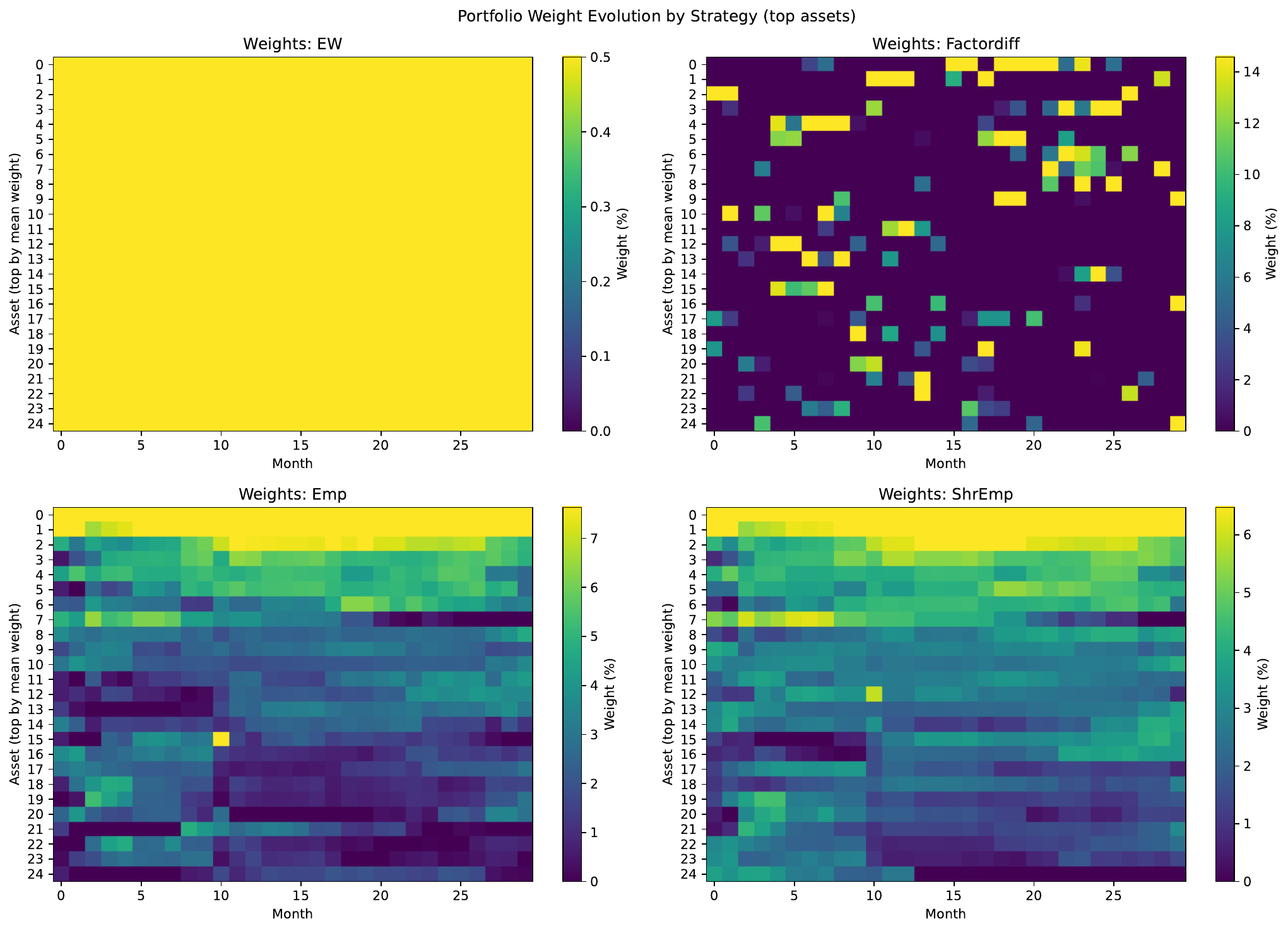}
\caption{Weights heatmap ($k=115$)}
\end{figure}

\begin{figure}[H]
\centering
\includegraphics[width=0.9\linewidth]{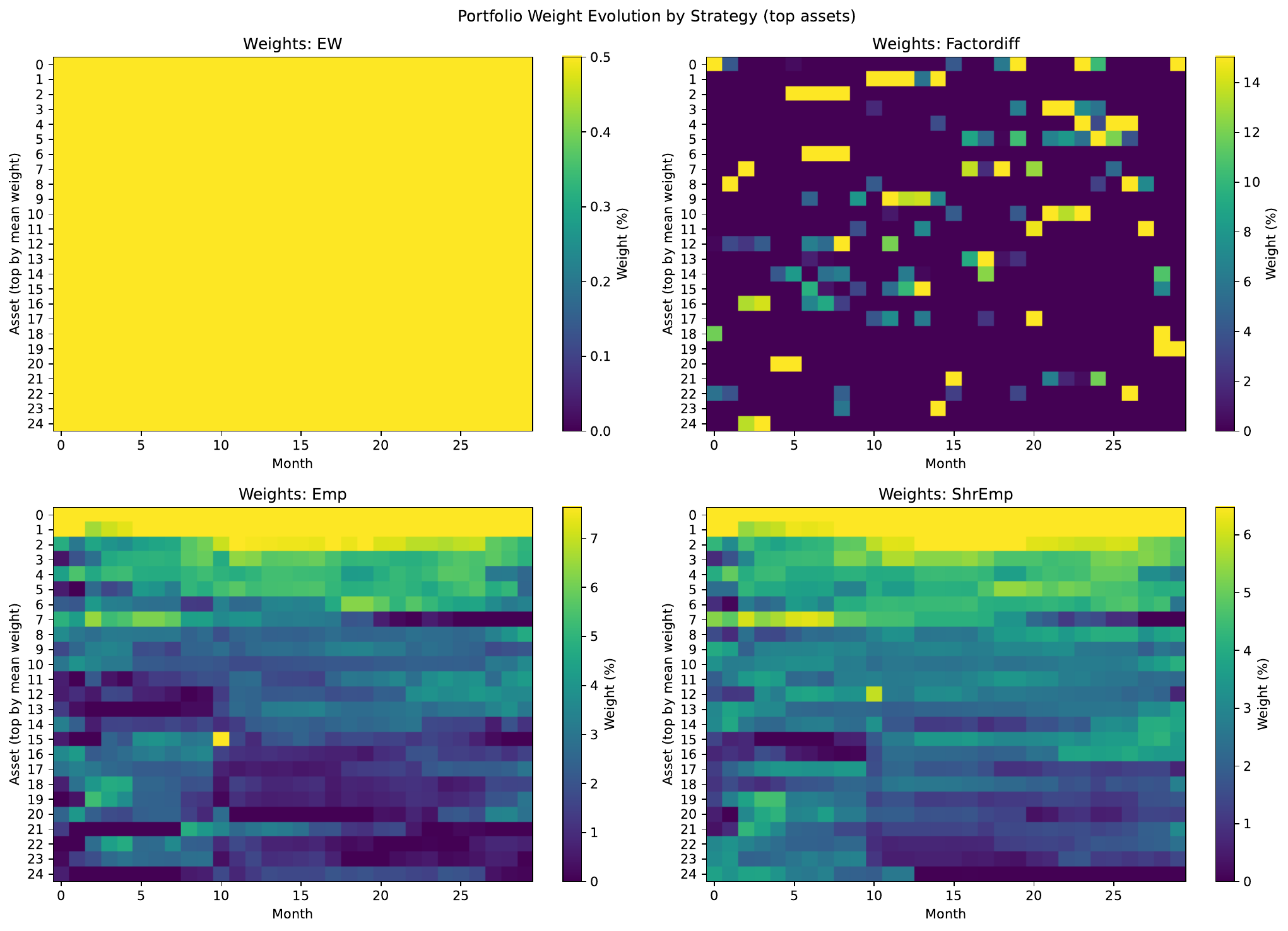}
\caption{Weights heatmap ($k=170$)}
\end{figure}

\begin{figure}[H]
\centering
\includegraphics[width=0.9\linewidth]{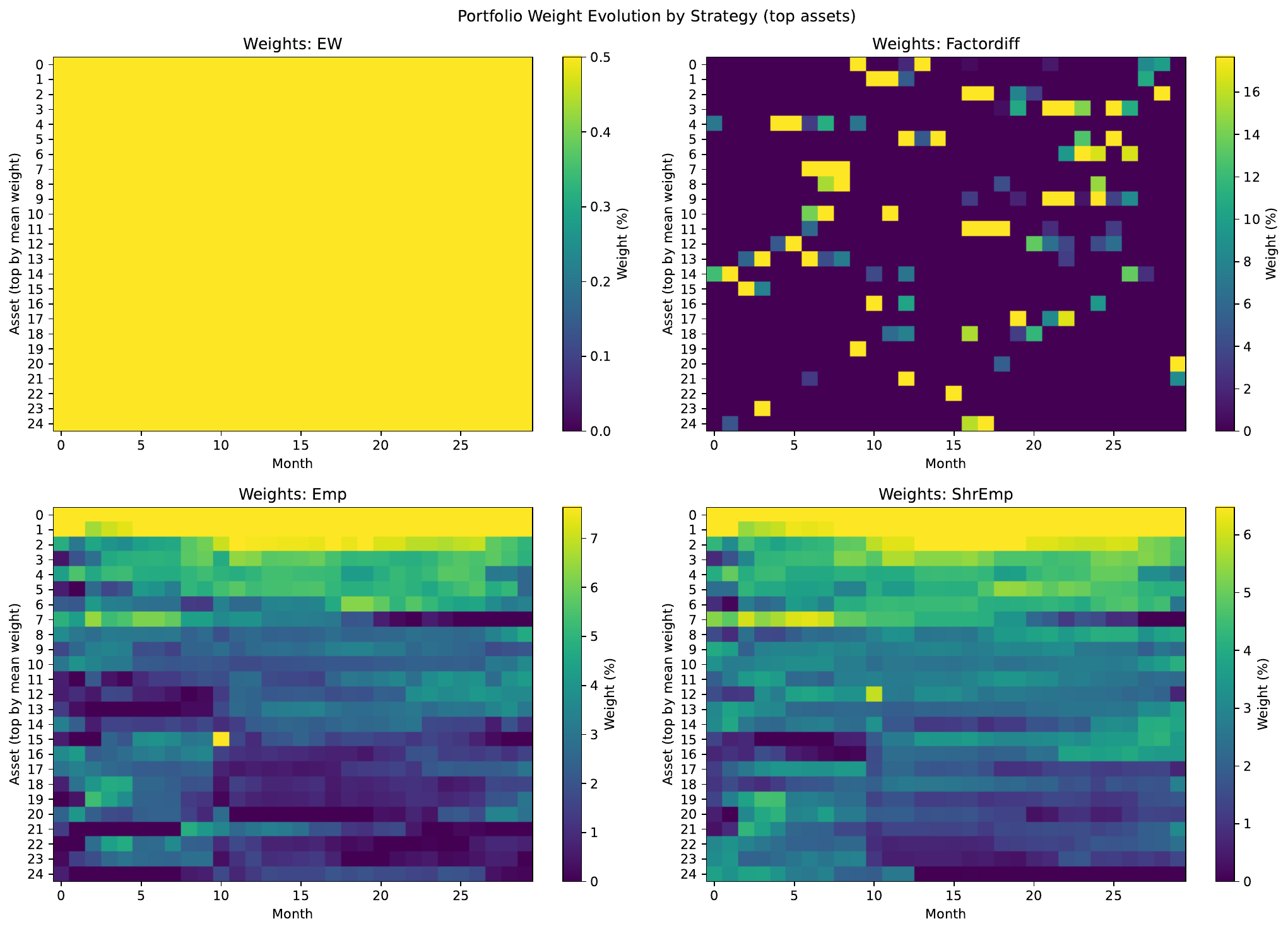}
\caption{Weights heatmap ($k=240$)}
\end{figure}

\begin{figure}[H]
\centering
\includegraphics[width=0.9\linewidth]{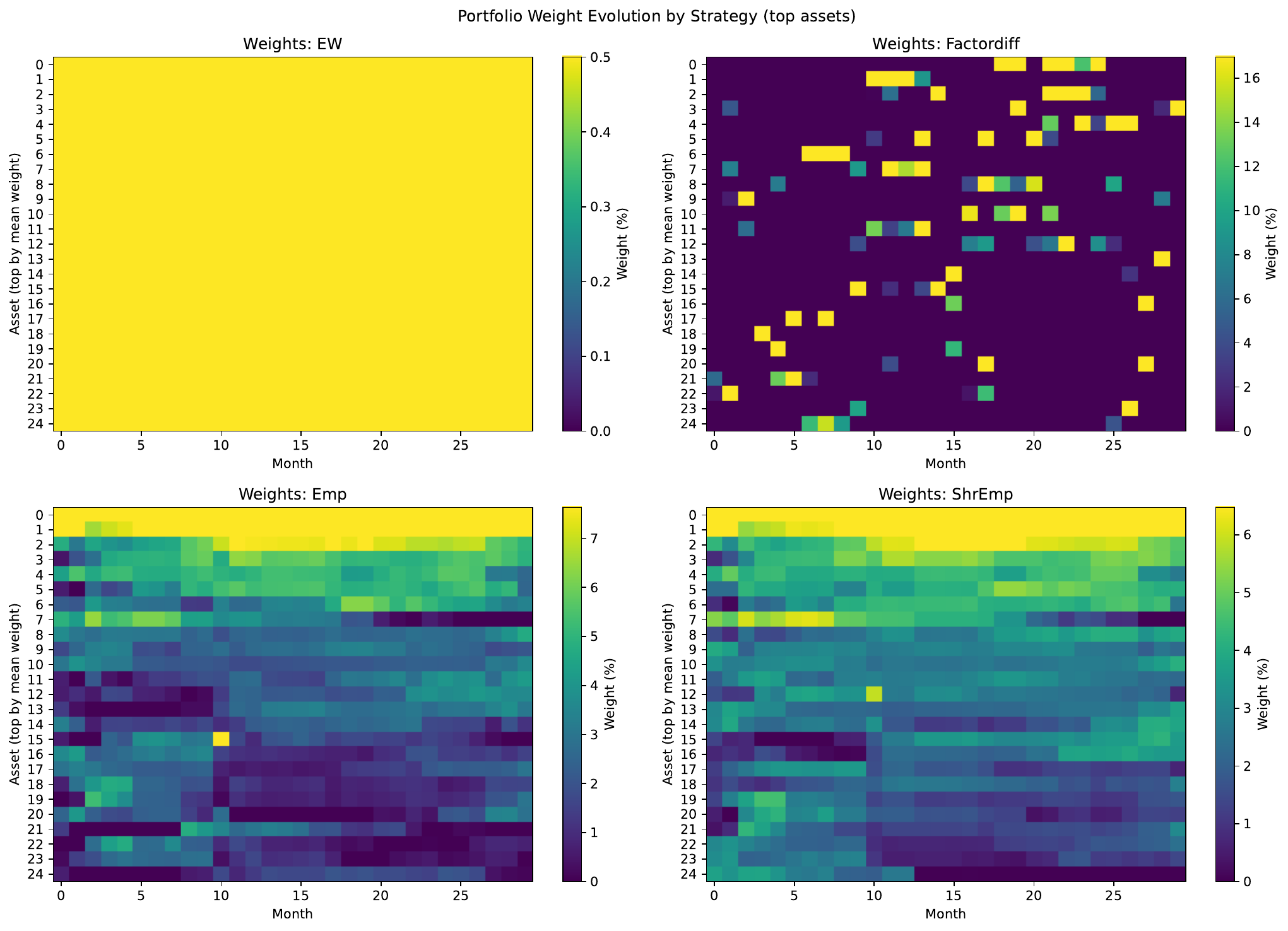}
\caption{Weights heatmap ($k=300$)}
\end{figure}

\begin{figure}[H]
\centering
\includegraphics[width=0.9\linewidth]{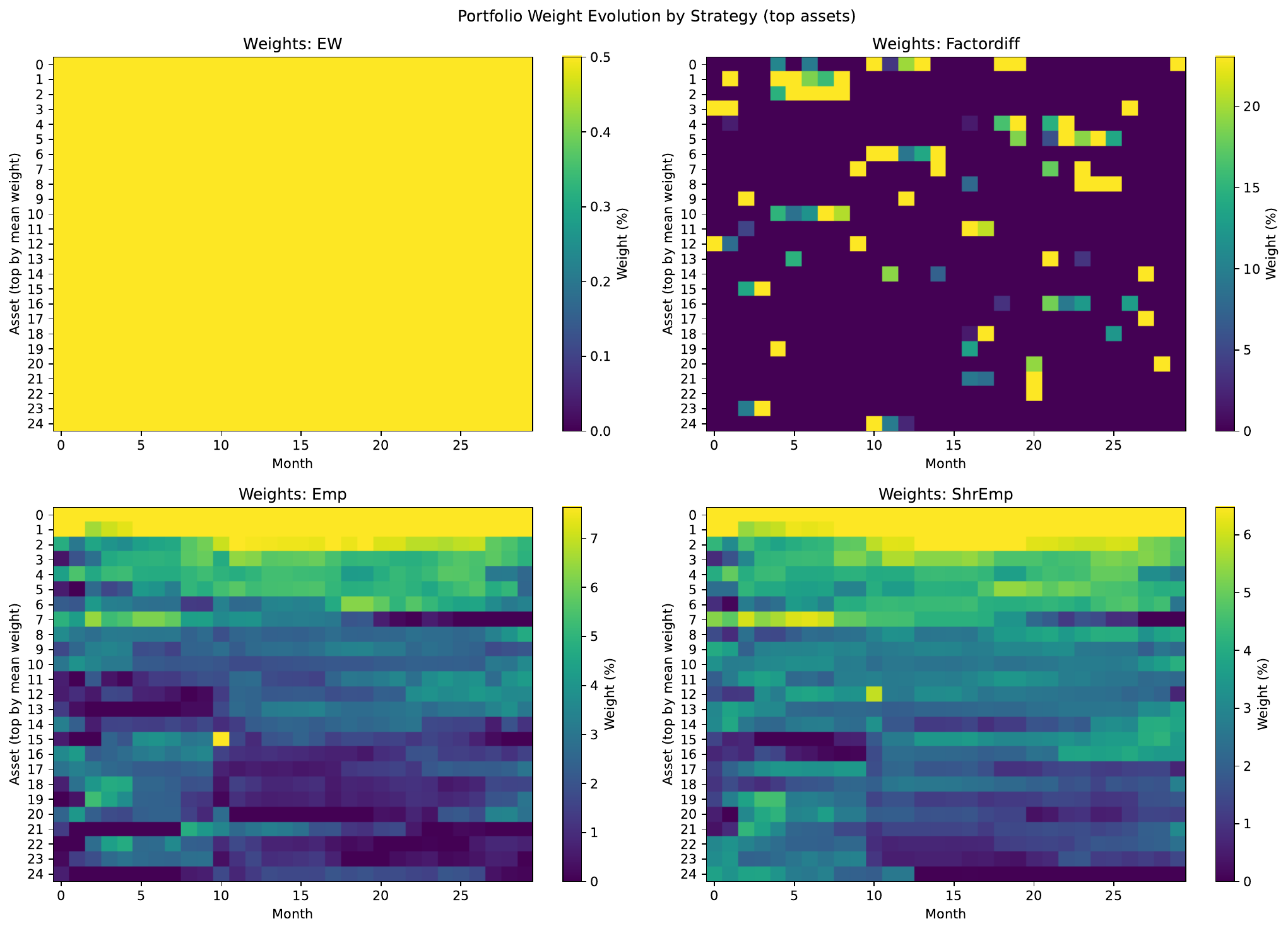}
\caption{Weights heatmap ($k=350$)}
\end{figure}

\subsection{Implicit Factor Modeling}\label{app:implicit}

\begin{figure}[h]
    \centering
    \includegraphics[width=0.6\textwidth]{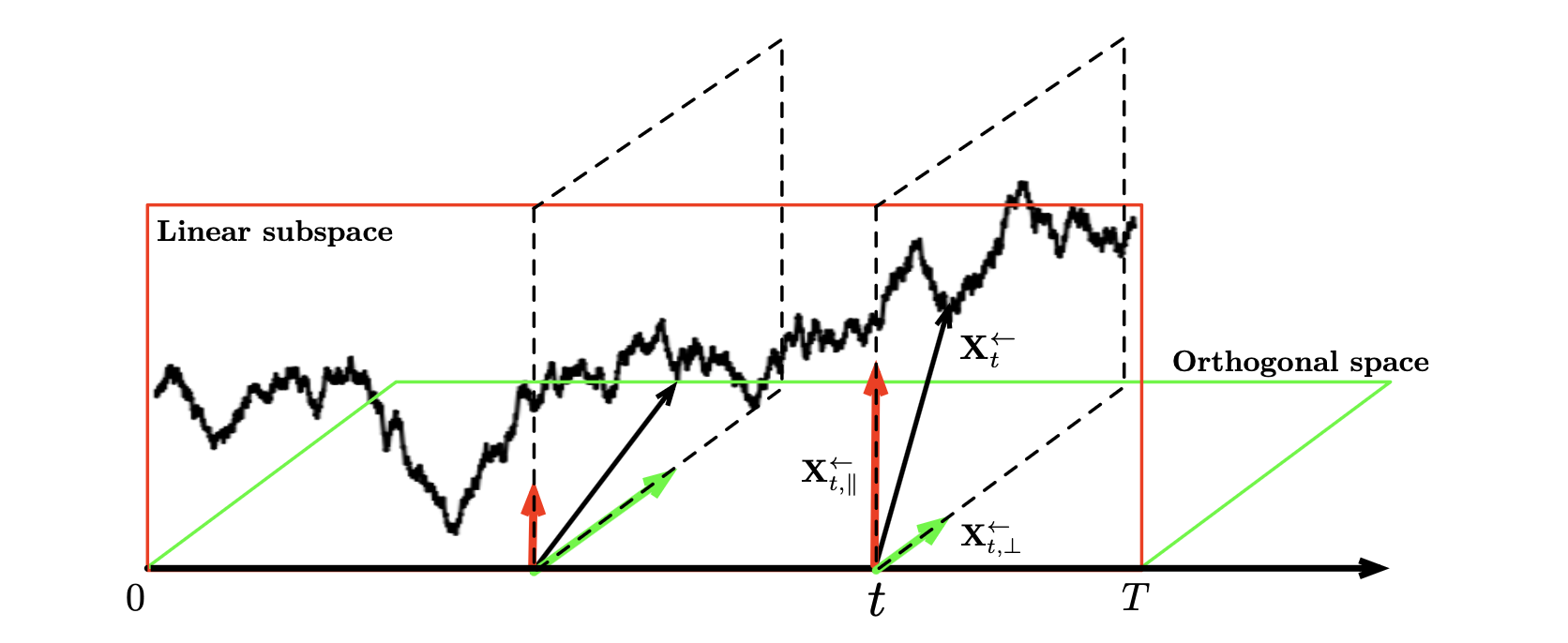}
    \caption{Image from \citet{chen2023scoreapproximationestimationdistribution}}
    \label{app:fig:linear_subspace}
\end{figure}

Figure~\ref{app:fig:linear_subspace} illustrates the decomposition proposed by \citet{chen2023scoreapproximationestimationdistribution}, where observed trajectories are separated into a low-dimensional linear subspace capturing systematic structure and an orthogonal component representing idiosyncratic variation. The projection onto the linear subspace can be interpreted as the evolution of factors, while the orthogonal space captures residual noise. Rather than explicitly specifying factors, score estimation learns this decomposition implicitly by modeling gradients of the data density along both directions. This perspective motivates implicit factor modeling, where diffusion models recover low-dimensional structure directly through score learning without requiring predefined factor exposures. Future studies should evaluate whether this approach can match the results following the factor dimension ablation in this work. 

\end{document}